\documentclass[aps,pra,groupedaddress,twocolumn,showpacs]{revtex4}

\usepackage{amsmath}
\usepackage{amssymb}
\usepackage{graphicx}	
\usepackage{bm} 
\usepackage{color}
\usepackage{epstopdf}

\DeclareMathOperator{\RealR}{\mathbb{R}}
\begin{document}
\title{Quantum single-particle properties in a one-dimensional curved space}

\author{J.~K. Pedersen}
\author{D.~V. Fedorov}
\author{A.~S. Jensen}
\author{N.~T. Zinner}
\affiliation{Department of Physics and Astronomy, Aarhus University, DK-8000 Aarhus C, Denmark}

\date{\today}

\begin{abstract}
We consider one particle confined to a deformed one-dimensional wire.
The quantum mechanical equivalent of the classical problem is not
uniquely defined.  We describe several possible hamiltonians and
corresponding solutions for a finite wire with fixed endpoints and
non-vanishing curvature.  We compute and compare the disparate
eigenvalues and eigenfunctions obtained from different quantization
prescriptions.  The JWKB approximation without potential leads
precisely to the square well spectrum and the coordinate dependent
stretched or compressed box related eigenfunctions.  The geometric
potential arising from an adiabatic expansion in terms of curvature is
at best only valid for very small curvature.
\end{abstract}
\maketitle

\section{Introduction}

Advances in optical trapping techniques makes it possible to trap cold atoms in what is effectively 1 or 2 dimensional\cite{ricardez2010,reitz2012,arnold2012,macdonald2002,bhatta2007,sague2008,pang2005}. 
An interesting direction of this is the ability to create effective quantum wires of varying shape and curvature.
An advantage of these lower dimensional setups are the how they can be shaped in ways that provide stabillity against short range collapse of polar molecules. Another thing is how long range interactions can shortcut through the second or third dimensions, and create novel interactions in 1 and 2 dimension
\cite{law2008,huhta2010,schmelcher2011,zampetaki2013,pedersen2014}.

When describing the motion of an atom moving along such a wire there
are at least two different approaches. The first is to solve the
problem fully in three dimensions, and treat the trapping potential as
any other potential. This approach is difficult to solve both
analytically and numerically, unless the trapping potential is highly
symmetric.  The other approach is to develop a quantum mechanical
description of the system in one dimension.  How to do this is not
well defined for general structures.  This paper will discuss various
quantization prescriptions of the classical motion, both in general
and in particular focussed on the one-dimensional wire problem.

In general, position dependent masses present problems in quantizing
classical motion.  This problem was already recognized by
Schr\"{o}dinger \cite{sch26}, and in the dynamical evolution of
nuclear shapes from equilibrium through saddle points to separated
fission fragments \cite{hof71}.  The problem appears first of all when
the starting points are generalized masses obtained independent from
potential energies as for example through classical cranking type of models
\cite{pau74} and micro-macro energy calculations \cite{bra72}

The motion can be constrained geometrically by an external field and
it is then possible to simulate this physical situation by steep
confining walls in forbidden directions.  Then the procedure can be
more straightforward by approximately solving the full
many-dimensional problem. The effects of the repulsive walls may then
be approximated by applying an adiabatic expansion which results in a
potential reflecting the geometric properties of the confinement
\cite{PhysRevA.89.033630}.  We shall compare results from different
quantization prescriptions.

The purpose of the present paper is to exhibit similarities and
differences between quantum mechanical results arising from different
quantizations of a single-particle hamiltonian in one dimension.  In
general we shall allow an undetermined coordinate dependent
parametrization of the particle mass.  In section \ref{sec:clas} we
develop a classical description of a particle confined to move in
along a curved wire.  In section \ref{sec:quantum} we show the
different choices of how to quantize the system. In the sections
\ref{sec:bulg} and \ref{sec:stretch} we look at three different
choices of wire, and using the different approaches from section
\ref{sec:quantum}, we calculate the spectra and the
eigenfunctions. Finally in section \ref{sec:outlook} we give a summary
of the discussions and an outlook.

\section{Classical Description\label{sec:clas}}

We consider one particle confined to move in a one-dimensional
trap. The particle is assumed to be point like with no internal
structure and have a mass of $m_0$. The trapping geometry is going to
be several different deformations of a helix. This one dimensional
curve is defined by a parametrization $F:\RealR\rightarrow\RealR^3$.
The parameterizations discussed in this paper will all be of the form
\begin{equation}
\left\{x,y,z\right\}=R\cdot\left\{f_x(\phi)\cos{\phi},f_y(\phi)\sin{\phi},f_z(\phi)\right\}.
\label{eq:xyz}
\end{equation}
Instead of using the full three dimensional set of coordinates we use
$\phi$, to describe the position of the particle along the curve.  

The classical velocity is then $(\dot{x},\dot{y},\dot{z})$, where the
dots denote time derivation.  The corresponding classical kinetic
energy, $T$, of the particle in three dimensions is then given by
\begin{equation}
T= \frac{1}{2}m_0\left(\dot{x}^2+\dot{y}^2+\dot{z}^2\right).
\end{equation}
Then $T$ can be transformed through Eq.~(\ref{eq:xyz}) to only depend
on the angular position, $\phi$, and velocity, $\dot{\phi}$, along the curve,
that is
\begin{eqnarray}
T(\phi,\dot{\phi}) = \frac{1}{2} m_0 (x'^2 + y'^2 + z'^2)  \dot{\phi}^2 
 \equiv \frac{1}{2}m(\phi)\dot{\phi}^2 \; , \label{eq:mass}
\end{eqnarray}
which defines the effective mass $m(\phi)$.  For the parametrization
in Eq.~(\ref{eq:xyz}) this yields the explicit expression
\begin{align}
\begin{split}
 m(\phi) = m_0 R^2 &\left[(f'^2_x+f^2_ y)\cos^2{\phi}+(f^2_x+f'^2_ y)
 \sin^2{\phi}\right. \\ &\left.
  + (f_xf'_x-f_yf'_y)\sin(2\phi) + f'^2_ z\right]  \; ,
\label{eq:mdef}
\end{split}
\end{align}
where the primes denote derivatives with respect to $\phi$.

A useful quantity is the canonical conjugate momentum, $p_{\phi}$, to the
position $\phi$ along the curve, that is \cite{taylor2005classical}
\begin{equation}
p_\phi=\frac{\partial T}{\partial\dot{\phi}}=m(\phi)\dot{\phi} \; ,
\end{equation} \label{eq:kin}
which provides the kinetic energy in canonical form
\begin{equation} \label{eq:kin1}
T=\frac{p_{\phi}^2}{2m(\phi)} \; .
\end{equation}
The coordinate dependent mass in Eq.~(\ref{eq:mdef}) is a key
quantity.  The present parametrization in Eq.~(\ref{eq:xyz}) only
provides $\phi$-independent effective mass when the functions $f'_z$
and $f_x = f_y$ are independent of $\phi$.  These are both necessary
and sufficient conditions for the constant effective mass,
$m(\phi)=m_0R^2(f_x^2+f'^2_z)$, of a regular helix.

Another simple case is $f_x = f_y$ where we have
\begin{equation}
m(\phi)=m_0R^2(f_x^2 + f'^2_x + f'^2_z),
\end{equation}
which still may contain a $\phi$-dependence.  When both $f_x$ and
$f_y$ are constants, which corresponds to the parametrization of an
ellipse in the x-y plane, we have 
\begin{equation}
m(\phi)=m_0R^2\left(f_x^2\cos^2{\phi}+f_y^2\sin^2{\phi}+f'^2_z\right).
\end{equation}
If furthermore, $f_x=f_y$ and $f'_z=0$ the helix reduces to a circle
in the $x-y$ plane, and the mass becomes $m_0R^2$, which is just the
initial mass scaled by the $R^2$-factor that appears when using the
dimensionless $\phi$-coordinate.  We conclude that the effects of a
non-constant effective mass require more than the regular helix, and
we need to consider angle-dependent $f$-functions.

\section{Quantizing the motion\label{sec:quantum}}

In this section we shall consider different prescriptions to quantize
the particle motion on the curves parameterized through the classical
physics described in the previous section.

\subsection{Quantizing in one dimension}

Without an external potential on the particle the hamiltonian only
contains terms arising from the kinetic energy in Eq.~(\ref{eq:kin1}).
The complication may be the position dependent effective mass.  The
straightforward quantization is from Eq.~(\ref{eq:kin1}) with
$p_{\phi}=-i \hbar \frac{\partial}{\partial\phi}$, but ordering of
this operator and the mass term, $1/m(\phi)$, is now important.  We
demand that the hamiltonian is hermitian and the eigenenergies of the
system must consequently be real.

A general structure of the kinetic energy operator is
\begin{equation}
 -\frac{\hbar^2}{2} \frac{1}{m^a}\frac{\partial}{\partial\phi}
 \frac{1}{m^b}\frac{\partial}{\partial\phi} \frac{1}{m^c} \; ,
\label{eq:eq26}
\end{equation}
where $a+b+c = 1$ and hermiticity requires $a=c$.  An even more
general form would be to allow $a \neq c$, but instead symmetrizing
afterwards, that is
\begin{equation}
 -\frac{\hbar^2}{4}\bigg( \frac{1}{m^a}\frac{\partial}{\partial\phi}
 \frac{1}{m^b}\frac{\partial}{\partial\phi} \frac{1}{m^c} +
 \frac{1}{m^c}\frac{\partial}{\partial\phi}
 \frac{1}{m^b}\frac{\partial}{\partial\phi} \frac{1}{m^a} \bigg) \; .
\label{eq:eq36}
\end{equation}
The quantization in Eq.~(\ref{eq:eq36}) is problematic for
discontinuous potentials, which breaks the hermitian property
\cite{dek99}.  This does not apply to our case and we shall not a
priori reject the two different terms in Eq.~(\ref{eq:eq36}).  To be
specific we choose typical examples with properties from
Eqs.~(\ref{eq:eq26}) and (\ref{eq:eq36}), that is $a=c=0, b=1$ and
$b=c=0, a=1$, respectively.

The first of these choices of an effective mass hamiltonian is with the mass
placed between the two momentum operators, that is
\begin{equation}
H_{1}=-\frac{\hbar^2}{2}\frac{\partial}{\partial\phi}\frac{1}{m(\phi)}\frac{\partial}{\partial\phi} \; .
\label{eq:pmp}
\end{equation}
By taking the first order derivative of the effective mass the
hamiltonian can be rewritten as
\begin{equation}
H_{1}=-\frac{\hbar^2}{2m(\phi)}\frac{\partial^2}{\partial\phi^2}+\frac{\hbar^2m'(\phi)}{2m^2(\phi)}\frac{\partial}{\partial\phi}.
\end{equation}
We note here that the position dependent effective mass, results in
an extra term in the hamiltonian.  Not only does this hamiltonian
contain a first order derivative of the effective mass, but this term 
also contains a first order derivative.  Both features are not
present in a normal hamiltonian only containing a kinetic energy part
with a constant mass.

The first order derivative can be removed as usual by the substitution
of the total wavefunction, $\Psi = m^{1/2}\psi$, into the
Schr\"{o}dinger equation, $(H_{1}-E)\Psi = 0$.  The corresponding new
hamiltonian acting on the reduced wavefunction, $\psi$ then becomes
\begin{equation} \label{eq:H1red}
 {H}_{EM_1}=-\frac{\hbar^2}{2m(\phi)}\frac{\partial^2}{\partial\phi^2}
 - \frac{\hbar^2}{4m^2(\phi)}  \bigg(m''(\phi) - \frac{3m'^2(\phi)}{2m(\phi)} 
 \bigg) \;,
\end{equation}
where the energy remains unchanged while a centrifugal barrier term
appears, and the Schr\"{o}dinger equation becomes $(H_{EM_1}-E) \psi =
0$.

Another possible choice of hamiltonian is one explicitly constructed
as hermitian with two terms.  One term with the effective mass before
the momentum operators, and another term with the effective mass after
the momentum operators, that is
\begin{equation} \label{eq:H1}
 H_{2} = -\frac{\hbar^2}{4}\left(\frac{\partial^2}{\partial\phi^2} \frac{1}{m(\phi)}
 + \frac{1}{m(\phi)} \frac{\partial^2}{\partial\phi^2} \right) \; .
\end{equation}
Calculating the derivatives of $m(\phi)$ we end up with a
hamiltonian, with the same first order derivative term appearing in $H_{1}$,
and in addition another effective potential. In total we get
\begin{equation}
H_{2}=-\frac{\hbar^2}{2m(\phi)}\frac{\partial^2}{\partial\phi^2}+\frac{\hbar^2m'(\phi)}{2m^2(\phi)}\frac{\partial}{\partial\phi}-\frac{\hbar^2}{4}
\frac{\partial^2}{\partial\phi^2}\left(\frac{1}{m(\phi)}\right).
\label{eq:H2}
\end{equation}
The extra potential term is the only difference between $H_{1}$ and
$H_{2}$. It contains terms of up to second order in the derivative of
the mass. The first order angular derivative is removed as in $H_{1}$
resulting in the hamiltonian, $H_{EM_2}$, acting on the reduced
wavefunction, $\psi$.  The difference between these two hamiltonians
amounts to an additional potential, that is
\begin{eqnarray}
V_{EM_2}&=&H_{EM_1}-H_{EM_2}  \label{eq:H1-H2}  \\ \nonumber 
&=& \frac{\hbar^2}{4}
\frac{\partial^2}{\partial\phi^2}\left(\frac{1}{m(\phi)}\right) 
 = \frac{\hbar^2}{4m^2}\big(m''-2\frac{m'^2}{m}\big) \;.
\end{eqnarray}
Thus, if the inverse of the effective mass by chance or by choice has
a vanishing second derivative, the potential, $V_{EM_2}$, in
Eq.~(\ref{eq:H1-H2}) is zero, and the two different schemes for
quantization yields exactly the same results.  

Specifically, this means that $H_{EM_1}=H_{EM_2}$ if a given
parametrization produces an effective mass such that
$1/m(\phi)=c_0+c_1\phi$, where $c_0$ and $c_1$ are constants.  We
shall later study a system with such an effective mass.  Extending to
a quadratic power dependence in $1/m(\phi)$ would produce a constant
difference between $H_{EM_1}$ and $H_{EM_2}$ without leading to any
difference in structure.  We emphasize that the two choices of
hamiltonian clearly only are examples.

\subsection{Quantizing in three dimensions}

A different approach to a quantum mechanical description of a curved
wire is found in \cite{PhysRevA.89.033630}.  They start with a
potential in three dimensions, assume a very tight confinement in the
directions perpendicular to the wire, and apply an adiabatic
expansion.  Only the lowest excited transverse state is allowed and a
one dimensional hamiltonian is then derived, that is
\begin{equation} \label{eq:geo}
H_{g}=-\frac{\hbar^2}{2m_0}\frac{\partial^2}{\partial s^2}-\frac{\hbar^2\kappa^2}{8m_0} \;,
\end{equation}
where $s$ is the arc length along the wire, and $\kappa$ is the
curvature of the wire as defined in the Frenet-Serret apparatus
\cite{docarmo1976}.  The last term in Eq.~(\ref{eq:geo}) is the
geometric potential, and the curvature for a curve in $\RealR^3$
parametrized by Eq.~(\ref{eq:xyz}) is explicitly given in
\cite{docarmo1976} to be
\begin{equation} \label{eq:kappa}
\kappa^2=\frac{\left(z''y'-y''z'\right)^2+\left(x''z'-z''x'\right)^2+\left(y''x'-x''y'\right)^2}{\left(x'^2+y'^2+z'^2\right)^3} ,
\end{equation}
where the primes again denote derivatives with respect to $\phi$.  

The geometric potential is always attractive, and most attractive
where the curvature is large.  The curve is here parameterized by the
arc length $s$, whereas we above used the azimuthal angle, $\phi$, to
specify the position.  The connection can be found by calculating $s$
as function of $\phi$, and if necessary invert the resulting
expression to get $\phi(s)$.

The arc length of a curve in $\RealR^3$ defined by Eq.~(\ref{eq:xyz})
can be calculated by \cite{docarmo1976}
\begin{equation}
 s(\phi)=\int_{\phi_{min}}^{\phi} d\varphi {\sqrt{x'^2+y'^2+z'^2}} =
\int_{\phi_{min}}^{\phi}{\sqrt{\frac{m(\varphi)}{m_0}}} d\varphi \; , \label{eq:arc}
\end{equation}
where we measured from $ \phi = {\phi_{min}}$ and used
Eq.~(\ref{eq:mass}) and the related coordinate dependence on the
angle, $\phi$.  This means that $s' = \sqrt{m(\phi)/m_0}$, which can
be used to transform the hamiltonian in Eq.~(\ref{eq:geo}) from the $s$ to
the $\phi$ coordinate.  The result is
\begin{align}
H_{g}&=-\frac{\hbar^2}{2m(\phi)}\frac{\partial^2}{\partial\phi^2}+\frac{\hbar^2m'(\phi)}{2m^2(\phi)}\frac{\partial}{\partial\phi}\nonumber \\
&+\frac{\hbar^2m''(\phi)}{8m^2(\phi)}-\frac{7\hbar^2m'^2(\phi)}{32m^3(\phi)}-\frac{\hbar^2\kappa^2}{8m_0} \; ,
\label{eq:fullHgeo}
\end{align} 
which can be verified most easily by going from
Eq.~(\ref{eq:fullHgeo}) to Eq.~(\ref{eq:geo}) by use of $\partial /
\partial \phi = \partial s/ \partial \phi \times \partial / \partial s$.

The geometric potential or rather the curvature, $\kappa$, can be
rewritten in terms of $m$, $m'$ and $m''$, and some extra terms
containing third derivatives with respect to $\phi$, that is
\begin{eqnarray} \label{eq:kappa2}
\kappa^2=\frac{\frac{1}{2}m m'' - \frac{1}{4}m'^2 - m (x'x'''+y'y'''+z'z''')}{m^3} \;,
\end{eqnarray}
which can be verified by use of Eqs.~(\ref{eq:mass}) and
(\ref{eq:kappa}).  Thus the curvature can be expressed through $m$,
$m'$, $m''$, and additional terms containing more than third
derivatives with respect to $\phi$.  Again we find the same first
order derivative term as in $H_1$ and $H_2$.  Removal results in the
hamiltonian, $H_{geo}$, acting on the reduced wavefunction.

We now have three different hamiltonians describing the same system.
They are all one-dimensional Schr\"{o}dinger equations with first and
second derivatives as well as various potential terms.  The first two,
$H_{EM_1}$ and $H_{EM_2}$, only have kinetic energy terms whereas the
last one, $H_{geo}$, also contains the attractive geometric potential.
However, depending on quantization prescription, the kinetic energy
parts differ from each other in these three cases, although all of
them have the ordinary second order derivative term with the same
coordinate dependent mass, that is
$-\frac{\hbar^2}{2m}\frac{\partial^2}{\partial\phi^2}$.  The same
first order derivative term,
$\frac{m'(\phi)}{4m^2(\phi)}\frac{\partial}{\partial\phi}$, has been
removed in all three cases.

The hamiltonian, $H_{geo}$, with the geometric potential differs
substantially from the other ones.  We note that the first two terms
in Eq.~(\ref{eq:kappa2}) containing derivatives of the mass are also
present in $H_{EM_2}$, except that they appear with different
strengths.  However, the last term in Eq.~(\ref{eq:kappa2}) has a
different structure with higher order derivatives of the coordinates
in three dimensions.

We emphasize that the geometric hamiltonian is derived under various
assumptions, and especially the adiabatic approximation which cannot
accommodate too rapid changes of the coordinates.  This means in
particular that at most terms up to second order derivatives are
correctly included, while derivative terms of third an higher order
have been neglected.  The accuracy of such approximations is dubious,
because higher order derivatives of the periodic trigonometric
functions in the parametrization are not decreasing in size.  This
does not prove that the result is inaccurate but more information is
necessary to evaluate the consequences of these assumptions.  

The curvature is an important quantity, at least as long as it remains
modest in size.  It is then of interest to know when it vanishes, or
equivalently when it becomes small.  From Eq.~(\ref{eq:kappa}) we see
that $\kappa = 0$ is obtained when $z''y'- y''z' = x''y' - x' y'' =
z'' x' -x'' z' = 0$. These differential equations can be rewritten
$(\ln x')' = (\ln y')' =(\ln z')'$ with the complete solutions $z
=ay+b=cx+d$ for arbitrary constants $a,b,c,d$.  Thus, a linear
dependence between all the coordinates eliminates the curvature.  This
is of course not surprising since a straight line by definition should
have curvature zero.  Still, it emphasizes the point that only a
modest correlated coordinate variation is allowed to maintain a small
curvature and a fairly accurate adiabatic expansion.  A
helix seems to be far away from this assumption.

\subsection{Semi-classical approach}

The search for an appropriate quantization prescription of a
classically well defined problem strongly suggest use of
semi-classical methods.  We therefore turn to the JWKB (Jeffreys-Wentzel-Kramers-Brillouin) approximation
which directly is applicable on a one-dimensional problem.
The lowest order expression for the wave function of a bound state is
\begin{align}
\begin{split}
 \Psi_{JWKB}(\phi)&=A\cos{\left[\int_{\phi_{min}}^{\phi} d\varphi
 \sqrt{\frac{2m(\varphi)}{\hbar^2}(E-V(\varphi))}d\varphi\right]} \\
 &+B\sin{\left[\int_{\phi_{min}}^{\phi} d\varphi \sqrt{\frac{2 m(\varphi)}
 {\hbar^2}(E-V(\varphi))}\right]} \; ,
\label{eq:wkbsin}
\end{split}
\end{align}
where $E$ is the energy of the particle, $A$ and $B$ are constants,
$\phi_{min}$ is one end of the wire, and $V(\phi)$ is the potential
along the wire.  The integral has to extend over all classically
allowed regions of $\phi$, that is where $E \ge V(\phi)$.  The
expression in the exponent is in fact found as an integral over the
classical canonical momentum, $p_\phi$, derived from
Eq.~(\ref{eq:kin1}) and the constraint of energy conservation $T=E-V$.

This choice is not unambiguous due to the coordinate dependence of the
effective mass.  We could choose one of the hamiltonians as the
starting point, then rewrite as in Eq.~(\ref{eq:H1red}) where the
first order derivative is removed and a reduced equation obtained.  This
is similar to the use of spherical coordinates and the equation for
the reduced radial wave function. Then the extra centrifugal potential
should be included.  We could also start with $H_{EM_2}$ and only include
the potential in Eq.~(\ref{eq:H1-H2}), or for that matter any linear
combination of these potentials. 

The simple JWKB wave function in Eq.~(\ref{eq:wkbsin}) could be
improved but the full solutions are easily available to us for the
different hamiltonians.  We shall therefore only use the JWKB
approximation to gain qualitative insight.  It is obvious that both
JWKB approximation and the geometric potential are only reliable for
modestly varying coordinates along the wire and in turn slowly varying
effective mass.  The different hamiltonians are then also rather
similar as their differences stem from the derivatives of mass or
coordinates. Thus, it suffice to use $V=0$ in Eq.~(\ref{eq:wkbsin}).

The boundary conditions we choose are precisely vanishing wave
function at the points terminating the classically allowed regions
as for example at the two ends of the finite wire.  This immediately
requires that $A=0$ since $\Psi_{JWKB}(\phi=\phi_{min})=0$.  The other
end point condition of $\Psi_{JWKB}(\phi=\phi_{max})=0$ then provides
the general quantization condition, that is
\begin{equation}\label{eq:wkbquant}
\sqrt{\frac{2E_n}{\hbar^2}}\int_{\phi_{min}}^{\phi_{max}}{\sqrt{m(\phi)}d\phi}=n\pi \;,
\end{equation} 
This equation is only fulfilled for discrete values of $E=E_n$, and
the corresponding wavefunctions are then given by
Eq.~(\ref{eq:wkbsin}) for $A=0$.

The integral over $\sqrt{m(\phi)}$ is measuring the total
length of the curve as seen by Eq~(\ref{eq:arc}).  The spectrum is
therefore exactly that of a particle in an infinitely deep square well, that is
\begin{eqnarray} \label{wkbenergy}
  E_n = \frac{\hbar^2 \pi^2 n^2}{2m_0  L^2(\phi_{max})} ,\;\;
    L(\phi) = \int_{\phi_{min}}^{\phi}{\sqrt{m(\phi)/m_0}d\phi} \;,
\end{eqnarray}  
where $L(\phi_{max})$ is the length of the one-dimensional box.
The corresponding eigenfunction is
\begin{equation} \label{wkbwave}
  \Psi_{JWKB} \propto \sin\big(n\pi L(\phi)/ (L(\phi_{max})\big) \; .
\end{equation}  

We can now use this simple expression to compute the JWKB wave function
for particular parameterizations where $\sqrt{m(\phi)}$ can be
analytically integrated.  Thus, the spectrum is that of a square well,
and the related eigenfunctions are deformed (stretched or contracted)
one-dimensional box wavefunctions.


\section{Bulging helix\label{sec:bulg}}
 
We shall now calculate the properties of the quantized structures.  We
must then first decide on an appropriate parametrization of the
one-dimensional curve.  Second we compare numerically the resulting
coordinate dependent masses, curvatures, spectra and eigenfunctions
for different quantization prescriptions.

\subsection{Parametrization}

\begin{figure}
\centering
\includegraphics[width=\columnwidth]{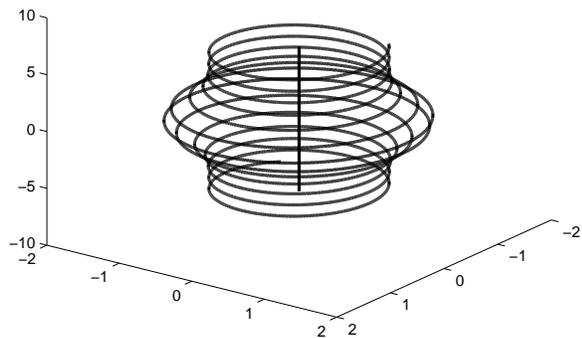}
\caption{A schematic drawing of a helix with a gaussian deformation of
  the type in Eq.(\ref{eq:gausdef}). The parameters are chosen to be
  $a=0.5$ and $\phi_0=4\phi$, and the radius then varies by a factor
  of $1.5$ over about $4$ windings.  The length can be extended as desired.
 \label{fig:gaushelix}}
\end{figure}

In section~\ref{sec:clas} we showed that a regular helix leads to a
constant effective mass.  The curvature is then naturally also small
or at least varying slowly with position along the one-dimensional
curve.  The different quantization prescriptions are then identical,
except perhaps for the one-dimensional adiabatic approximation of
the three-dimensional result.  To emphasize the effects of varying
mass and curvature we therefore start by modifying the simple helix.
We keep the circular nature, that is the radial variation of the $x$
and $y$-directions are chosen to be identical, but varying quickly
over a few windings.  The unrelated $z$-direction is linear to
maintain the equidistant helix structure.  In total, we choose the
parametrization in Eq.~(\ref{eq:xyz}) to be
\begin{align}
\begin{split}
f_x(\phi)&=f_y(\phi)=\left(1+a\cdot\exp(-\phi^2/\phi_0^2)\right).
\label{eq:gausdef}\\
f_z(\phi)&=\phi  \;,
\end{split}
\end{align} 
where $a$ and $\phi_0$ are constants.  For practical convenience we
use the gaussian to describe the form of the variation.  The size of
$a$ determines the radial changes, ranging from $R$ to $R(1+a)$ and
back again as $\phi$ varies from $-\infty$ to $+\infty$.  The width of
the gaussian, $\phi_0$, determines how quickly the radial change is
taking place, that is over how many windings.  The length of the curve
is chosen to be from $\phi_{min}=-40$ to $\phi_{max}=40$, which is
sufficient to allow a bump extending over several windings while the
radius return to the initial value at the end points, that is when
$f_x(\phi)=f_{y}(\phi)\approx R$.  In numbers, $\phi_0 = 2\pi$ means
very fast variation over one winding.  A given multiplum, $\ell$, of
$2\pi$ then implies the gaussian variation over $2 \ell$ windings.
The geometric structure is illustrated schematically in
fig.~\ref{fig:gaushelix}.

\begin{figure}
\centering
\begingroup
  \makeatletter
  \providecommand\color[2][]{%
    \GenericError{(gnuplot) \space\space\space\@spaces}{%
      Package color not loaded in conjunction with
      terminal option `colourtext'%
    }{See the gnuplot documentation for explanation.%
    }{Either use 'blacktext' in gnuplot or load the package
      color.sty in LaTeX.}%
    \renewcommand\color[2][]{}%
  }%
  \providecommand\includegraphics[2][]{%
    \GenericError{(gnuplot) \space\space\space\@spaces}{%
      Package graphicx or graphics not loaded%
    }{See the gnuplot documentation for explanation.%
    }{The gnuplot epslatex terminal needs graphicx.sty or graphics.sty.}%
    \renewcommand\includegraphics[2][]{}%
  }%
  \providecommand\rotatebox[2]{#2}%
  \@ifundefined{ifGPcolor}{%
    \newif\ifGPcolor
    \GPcolortrue
  }{}%
  \@ifundefined{ifGPblacktext}{%
    \newif\ifGPblacktext
    \GPblacktextfalse
  }{}%
  \let\gplgaddtomacro\g@addto@macro
  \gdef\gplbacktext{}%
  \gdef\gplfronttext{}%
  \makeatother
  \ifGPblacktext
    \def\colorrgb#1{}%
    \def\colorgray#1{}%
  \else
    \ifGPcolor
      \def\colorrgb#1{\color[rgb]{#1}}%
      \def\colorgray#1{\color[gray]{#1}}%
      \expandafter\def\csname LTw\endcsname{\color{white}}%
      \expandafter\def\csname LTb\endcsname{\color{black}}%
      \expandafter\def\csname LTa\endcsname{\color{black}}%
      \expandafter\def\csname LT0\endcsname{\color[rgb]{1,0,0}}%
      \expandafter\def\csname LT1\endcsname{\color[rgb]{0,1,0}}%
      \expandafter\def\csname LT2\endcsname{\color[rgb]{0,0,1}}%
      \expandafter\def\csname LT3\endcsname{\color[rgb]{1,0,1}}%
      \expandafter\def\csname LT4\endcsname{\color[rgb]{0,1,1}}%
      \expandafter\def\csname LT5\endcsname{\color[rgb]{1,1,0}}%
      \expandafter\def\csname LT6\endcsname{\color[rgb]{0,0,0}}%
      \expandafter\def\csname LT7\endcsname{\color[rgb]{1,0.3,0}}%
      \expandafter\def\csname LT8\endcsname{\color[rgb]{0.5,0.5,0.5}}%
    \else
      \def\colorrgb#1{\color{black}}%
      \def\colorgray#1{\color[gray]{#1}}%
      \expandafter\def\csname LTw\endcsname{\color{white}}%
      \expandafter\def\csname LTb\endcsname{\color{black}}%
      \expandafter\def\csname LTa\endcsname{\color{black}}%
      \expandafter\def\csname LT0\endcsname{\color{black}}%
      \expandafter\def\csname LT1\endcsname{\color{black}}%
      \expandafter\def\csname LT2\endcsname{\color{black}}%
      \expandafter\def\csname LT3\endcsname{\color{black}}%
      \expandafter\def\csname LT4\endcsname{\color{black}}%
      \expandafter\def\csname LT5\endcsname{\color{black}}%
      \expandafter\def\csname LT6\endcsname{\color{black}}%
      \expandafter\def\csname LT7\endcsname{\color{black}}%
      \expandafter\def\csname LT8\endcsname{\color{black}}%
    \fi
  \fi
  \setlength{\unitlength}{0.0500bp}%
  \begin{picture}(3968.00,3968.00)%
    \gplgaddtomacro\gplbacktext{%
      \csname LTb\endcsname%
      \put(66,2067){\makebox(0,0)[r]{\strut{}-2}}%
      \put(66,2304){\makebox(0,0)[r]{\strut{} 0}}%
      \put(66,2541){\makebox(0,0)[r]{\strut{} 2}}%
      \put(66,2778){\makebox(0,0)[r]{\strut{} 4}}%
      \put(66,3015){\makebox(0,0)[r]{\strut{} 6}}%
      \put(66,3252){\makebox(0,0)[r]{\strut{} 8}}%
    }%
    \gplgaddtomacro\gplfronttext{%
      \csname LTb\endcsname%
      \put(2781,3198){\makebox(0,0)[r]{\strut{}$m(x)$}}%
      \csname LTb\endcsname%
      \put(2781,2978){\makebox(0,0)[r]{\strut{}$10m^,(x)$}}%
    }%
    \gplgaddtomacro\gplbacktext{%
      \csname LTb\endcsname%
      \put(66,2067){\makebox(0,0)[r]{\strut{}-2}}%
      \put(66,2304){\makebox(0,0)[r]{\strut{} 0}}%
      \put(66,2541){\makebox(0,0)[r]{\strut{} 2}}%
      \put(66,2778){\makebox(0,0)[r]{\strut{} 4}}%
      \put(66,3015){\makebox(0,0)[r]{\strut{} 6}}%
      \put(66,3252){\makebox(0,0)[r]{\strut{} 8}}%
    }%
    \gplgaddtomacro\gplfronttext{%
      \csname LTb\endcsname%
      \put(1386,3198){\makebox(0,0)[r]{\strut{}$50m^{,,}(x)$}}%
    }%
    \gplgaddtomacro\gplbacktext{%
      \csname LTb\endcsname%
      \put(66,595){\makebox(0,0)[r]{\strut{}-0.01}}%
      \put(66,826){\makebox(0,0)[r]{\strut{} 0}}%
      \put(66,1058){\makebox(0,0)[r]{\strut{} 0.01}}%
      \put(66,1289){\makebox(0,0)[r]{\strut{} 0.02}}%
      \put(66,1520){\makebox(0,0)[r]{\strut{} 0.03}}%
      \put(66,1752){\makebox(0,0)[r]{\strut{} 0.04}}%
      \put(198,375){\makebox(0,0){\strut{}-40}}%
      \put(644,375){\makebox(0,0){\strut{}-30}}%
      \put(1091,375){\makebox(0,0){\strut{}-20}}%
      \put(1537,375){\makebox(0,0){\strut{}-10}}%
      \put(1983,375){\makebox(0,0){\strut{} 0}}%
      \put(2429,375){\makebox(0,0){\strut{} 10}}%
      \put(2876,375){\makebox(0,0){\strut{} 20}}%
      \put(3322,375){\makebox(0,0){\strut{} 30}}%
      \put(3768,375){\makebox(0,0){\strut{} 40}}%
      \put(1983,45){\makebox(0,0){\strut{}$\phi$}}%
    }%
    \gplgaddtomacro\gplfronttext{%
      \csname LTb\endcsname%
      \put(858,1810){\makebox(0,0)[r]{\strut{}$V_{EM_2}$}}%
      \csname LTb\endcsname%
      \put(858,1590){\makebox(0,0)[r]{\strut{}$V_{geo}$}}%
      \csname LTb\endcsname%
      \put(858,1370){\makebox(0,0)[r]{\strut{}$V_{EM_{1}}$}}%
    }%
    \gplbacktext
    \put(0,0){\includegraphics{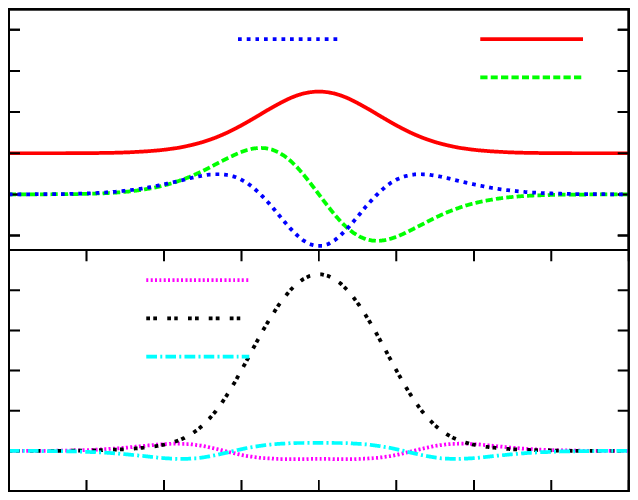}}%
    \gplfronttext
  \end{picture}%
\endgroup
\caption{Mass and corresponding derivatives as function of angle,
  $\phi$, for the parametrization in Eq.(\ref{eq:gausdef}).  Top:
  $m(\phi)$ (red), $10\cdot m'(\phi)$ (green), $10\cdot m''(\phi)$ (blue), all in
  units of $m_0R^2$.  Bottom contains potentials in units of
  $\frac{\hbar^2}{4m_0R^2}$ that enter different quantization
  prescriptions: $-\big(m''-
  \frac{3}{2}\frac{m'^2}{m}\big)/m^2$(magenta), $\big(m''-2\frac{m'^2}{m}\big)/m^2$(cyan), $-\kappa^2/2+0.1$(black), see
  Eqs.(\ref{eq:H1red}), (\ref{eq:H1-H2}) and (\ref{eq:geo}).
 \label{fig:gausmas}}
\end{figure}

The parametrization in Eq.(\ref{eq:gausdef}) leads immediately from
Eq.(\ref{eq:mdef}) to an effective mass given by
\begin{equation} \label{massbulge}
 m(\phi)=m_0R^2\left(1+\left(\frac{\partial f_x}
{\partial\phi}\right)^2+f_x^2\right) \;.
\end{equation}  
The first and second order derivatives of the mass are then easily
written down along with the potentials entering the expressions for
the different quantizations, that is Eqs.(\ref{eq:H1red}),
(\ref{eq:H1-H2}) and (\ref{eq:geo}).  We show their angular dependence
in fig.~\ref{fig:gausmas}.

The mass is constant for angles far away from the bulge on the
circular helix.  It exhibits a gaussian peak at the center with a width
of $2 \phi_0$ in accordance with Eq.(\ref{massbulge}).  This is then
the origin of a non-constant mass on the quantization, or equivalently
the effect from quantal motion in a one-dimensional curved space.

The size of the effects is reflected in the variation of the
derivatives shown in the same figure.  Both first and second
derivatives are rather small implying that the different quantizations
at least qualitatively should reveal the same most important features.
This is in spite of the initial periodic parameterization where all such
traces have disappeared since $f_x=f_y$ and the square of the sine and
cosine functions are added with equal amplitude.

The differences in quantization are quantified by the three
potentials at the bottom of fig.~\ref{fig:gausmas}. For $H_{EM_1}$ we have
the potential appearing in Eq.(\ref{eq:H1red}) after removal of the first
order operator derivative.  This term has an attractive minimum at the
bulge and symmetric repulsive maxima in the tail of the bulge.  The
difference between $H_{EM_1}$ and $H_{EM_2}$ is comparable in size and almost
flat over the bulge with two symmetric extremum points in the tail.

The transverse-mode adiabatic approximation quantization from
Eq.(\ref{eq:fullHgeo}) has two additional potentials compared to
$H_{EM_1}$, that is one very similar to Eq.(\ref{eq:H1-H2}) and the
qualitatively different geometric potential proportional to the square
of the curvature.  The latter is always attractive and of much larger
magnitude, but otherwise with the same behavior as the mass itself,
that is strongest in the center and vanishing outside the bulge, see
fig.~\ref{fig:gausmas}.

\subsection{Spectra}

We now solve the schr\"odinger equation numerically for the given
parametrization for all the different choices of quantization
described in section \ref{sec:quantum}.  We maintain the boundary
conditions corresponding to a curve of finite length and fixed
end-points, that $\Psi(\phi_{min})=\Psi(\phi_{max})=0$.  We discretize
the angular space by choosing a grid, computing the finite element
representation of the operators, and diagonalizing in the
corresponding basis.  We then compare eigenvalues and eigenfunctions
from the different quantization prescriptions.

\begin{figure}
\centering
\begingroup
  \makeatletter
  \providecommand\color[2][]{%
    \GenericError{(gnuplot) \space\space\space\@spaces}{%
      Package color not loaded in conjunction with
      terminal option `colourtext'%
    }{See the gnuplot documentation for explanation.%
    }{Either use 'blacktext' in gnuplot or load the package
      color.sty in LaTeX.}%
    \renewcommand\color[2][]{}%
  }%
  \providecommand\includegraphics[2][]{%
    \GenericError{(gnuplot) \space\space\space\@spaces}{%
      Package graphicx or graphics not loaded%
    }{See the gnuplot documentation for explanation.%
    }{The gnuplot epslatex terminal needs graphicx.sty or graphics.sty.}%
    \renewcommand\includegraphics[2][]{}%
  }%
  \providecommand\rotatebox[2]{#2}%
  \@ifundefined{ifGPcolor}{%
    \newif\ifGPcolor
    \GPcolortrue
  }{}%
  \@ifundefined{ifGPblacktext}{%
    \newif\ifGPblacktext
    \GPblacktextfalse
  }{}%
  \let\gplgaddtomacro\g@addto@macro
  \gdef\gplbacktext{}%
  \gdef\gplfronttext{}%
  \makeatother
  \ifGPblacktext
    \def\colorrgb#1{}%
    \def\colorgray#1{}%
  \else
    \ifGPcolor
      \def\colorrgb#1{\color[rgb]{#1}}%
      \def\colorgray#1{\color[gray]{#1}}%
      \expandafter\def\csname LTw\endcsname{\color{white}}%
      \expandafter\def\csname LTb\endcsname{\color{black}}%
      \expandafter\def\csname LTa\endcsname{\color{black}}%
      \expandafter\def\csname LT0\endcsname{\color[rgb]{1,0,0}}%
      \expandafter\def\csname LT1\endcsname{\color[rgb]{0,1,0}}%
      \expandafter\def\csname LT2\endcsname{\color[rgb]{0,0,1}}%
      \expandafter\def\csname LT3\endcsname{\color[rgb]{1,0,1}}%
      \expandafter\def\csname LT4\endcsname{\color[rgb]{0,1,1}}%
      \expandafter\def\csname LT5\endcsname{\color[rgb]{1,1,0}}%
      \expandafter\def\csname LT6\endcsname{\color[rgb]{0,0,0}}%
      \expandafter\def\csname LT7\endcsname{\color[rgb]{1,0.3,0}}%
      \expandafter\def\csname LT8\endcsname{\color[rgb]{0.5,0.5,0.5}}%
    \else
      \def\colorrgb#1{\color{black}}%
      \def\colorgray#1{\color[gray]{#1}}%
      \expandafter\def\csname LTw\endcsname{\color{white}}%
      \expandafter\def\csname LTb\endcsname{\color{black}}%
      \expandafter\def\csname LTa\endcsname{\color{black}}%
      \expandafter\def\csname LT0\endcsname{\color{black}}%
      \expandafter\def\csname LT1\endcsname{\color{black}}%
      \expandafter\def\csname LT2\endcsname{\color{black}}%
      \expandafter\def\csname LT3\endcsname{\color{black}}%
      \expandafter\def\csname LT4\endcsname{\color{black}}%
      \expandafter\def\csname LT5\endcsname{\color{black}}%
      \expandafter\def\csname LT6\endcsname{\color{black}}%
      \expandafter\def\csname LT7\endcsname{\color{black}}%
      \expandafter\def\csname LT8\endcsname{\color{black}}%
    \fi
  \fi
  \setlength{\unitlength}{0.0500bp}%
  \begin{picture}(3400.00,3968.00)%
    \gplgaddtomacro\gplbacktext{%
      \csname LTb\endcsname%
      \put(38,2103){\makebox(0,0)[r]{\strut{} 0}}%
      \put(38,2420){\makebox(0,0)[r]{\strut{} 0.002}}%
      \put(38,2737){\makebox(0,0)[r]{\strut{} 0.004}}%
      \put(38,3054){\makebox(0,0)[r]{\strut{} 0.006}}%
      \put(38,3371){\makebox(0,0)[r]{\strut{} 0.008}}%
    }%
    \gplgaddtomacro\gplfronttext{%
      \csname LTb\endcsname%
      \put(830,3198){\makebox(0,0)[r]{\strut{}$H_{EM_2}$}}%
      \csname LTb\endcsname%
      \put(2213,3198){\makebox(0,0)[r]{\strut{}$H_{EM_1}$}}%
    }%
    \gplgaddtomacro\gplbacktext{%
      \csname LTb\endcsname%
      \put(38,2103){\makebox(0,0)[r]{\strut{} 0}}%
      \put(38,2420){\makebox(0,0)[r]{\strut{} 0.002}}%
      \put(38,2737){\makebox(0,0)[r]{\strut{} 0.004}}%
      \put(38,3054){\makebox(0,0)[r]{\strut{} 0.006}}%
      \put(38,3371){\makebox(0,0)[r]{\strut{} 0.008}}%
    }%
    \gplgaddtomacro\gplfronttext{%
    }%
    \gplgaddtomacro\gplbacktext{%
      \csname LTb\endcsname%
      \put(38,595){\makebox(0,0)[r]{\strut{}-0.032}}%
      \put(38,849){\makebox(0,0)[r]{\strut{}-0.030}}%
      \put(38,1103){\makebox(0,0)[r]{\strut{}-0.028}}%
      \put(38,1356){\makebox(0,0)[r]{\strut{}-0.026}}%
      \put(38,1610){\makebox(0,0)[r]{\strut{}-0.024}}%
      \put(38,1864){\makebox(0,0)[r]{\strut{}-0.022}}%
      \put(1699,45){\makebox(0,0){\strut{}$a$}}%
    }%
    \gplgaddtomacro\gplfronttext{%
      \csname LTb\endcsname%
      \put(830,1691){\makebox(0,0)[r]{\strut{}$H_{geo}$}}%
    }%
    \gplgaddtomacro\gplbacktext{%
      \csname LTb\endcsname%
      \put(38,595){\makebox(0,0)[r]{\strut{}-0.032}}%
      \put(38,849){\makebox(0,0)[r]{\strut{}-0.030}}%
      \put(38,1103){\makebox(0,0)[r]{\strut{}-0.028}}%
      \put(38,1356){\makebox(0,0)[r]{\strut{}-0.026}}%
      \put(38,1610){\makebox(0,0)[r]{\strut{}-0.024}}%
      \put(38,1864){\makebox(0,0)[r]{\strut{}-0.022}}%
      \put(170,375){\makebox(0,0){\strut{} 0}}%
      \put(782,375){\makebox(0,0){\strut{} 0.2}}%
      \put(1394,375){\makebox(0,0){\strut{} 0.4}}%
      \put(2005,375){\makebox(0,0){\strut{} 0.6}}%
      \put(2617,375){\makebox(0,0){\strut{} 0.8}}%
      \put(3229,375){\makebox(0,0){\strut{} 1}}%
      \put(1699,45){\makebox(0,0){\strut{}$a$}}%
    }%
    \gplgaddtomacro\gplfronttext{%
    }%
    \gplbacktext
    \put(0,0){\includegraphics{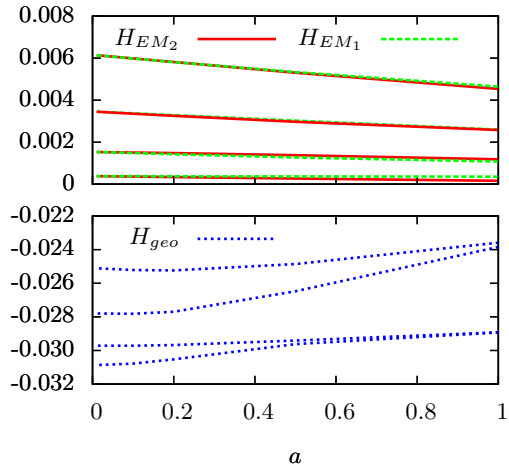}}%
    \gplfronttext
  \end{picture}%
\endgroup
\caption{The absolute energies of the four lowest states of one particle on a wire parametrized in Eq.~(\ref{eq:gausdef}) with $a=1$ and $\phi_0=4\pi$. The energies are in units of $\hbar^2/(R^2
  m_0)$. The upper panel shows the energies of $H_{EM_1}$ and $H_{EM_2}$ where as the lower one shows the energies of $H_{geo}$
 \label{fig:gausEofa}}
\end{figure}

\begin{table*}
\caption{The lowest four eigenvalues for different hamiltonians
  describing one particle on a wire parametrized in
  Eq.~(\ref{eq:gausdef}) with $a=1$ and $\phi_0=4\pi$.  The ground
  state energies are in absolute values in units of $\hbar^2/(R^2
  m_0)$.  The excited states are reported as excitation energies
  related to the corresponding ground state energy.
\label{tab:energiesa1}.}
\begin{ruledtabular}
 \begin{tabular}{c| c c c c c}
Hamiltonian		&	$H_{EM_1}$			&	$H_{EM_2}$ 			& JWKB 				&	$H_{geo}$		 &JWKB (including $V_{geo}$) \\
\hline
Ground state	&$3.58\cdot10^{-4}$	&$1.63\cdot10^{-4}$	&$2.88\cdot10^{-4}$ &$-2.89\cdot10^{-2}$ &$-2.76\cdot10^{-2}$ \\ \hline
1st excited		&$7.18\cdot10^{-4}$	&$1.02\cdot10^{-3}$	&$8.64\cdot10^{-4}$ &$8.62\cdot10^{-6}$ &$0$ \\
2nd excited		&$2.24\cdot10^{-3}$	&$2.42\cdot10^{-3}$	&$2.30\cdot10^{-3}$ &$5.07\cdot10^{-3}$ &$7.10\cdot10^{-3}$ \\ 
3rd excited		&$4.29\cdot10^{-3}$ &$4.37\cdot10^{-3}$	&$4.32\cdot10^{-3}$ &$5.34\cdot10^{-3}$ &$7.10\cdot10^{-3}$ \\
\end{tabular}
 \end{ruledtabular}
\end{table*}

\begin{table*}
 \caption{The same as in table~\ref{tab:energiesa1} for $a=0.5$ and
  $\phi_0=4\pi$.  
\label{tab:energiesa05}.}
\begin{ruledtabular}
 \begin{tabular}{c| c c c c c}
Hamiltonian		&	$H_{EM_1}$			&	$H_{EM_2}$ 			& JWKB 				&	$H_{geo}$		 &JWKB (including $V_{geo}$)\\
\hline
Ground state	&$3.72\cdot10^{-4}$	&$2.73\cdot10^{-4}$	&$3.34\cdot10^{-4}$ &$-2.96\cdot10^{-2}$ &$-2.86\cdot10^{-2}$ \\ \hline
1st excited		&$9.01\cdot10^{-4}$	&$1.10\cdot10^{-3}$	&$1.00\cdot10^{-3}$ &$2.07\cdot10^{-4}$ &$0$ \\
2nd excited		&$2.65\cdot10^{-3}$	&$2.70\cdot10^{-3}$	&$2.67\cdot10^{-3}$ &$3.14\cdot10^{-3}$ &$1.70\cdot10^{-3}$ \\ 
3rd excited		&$4.97\cdot10^{-3}$ &$5.03\cdot10^{-3}$	&$5.01\cdot10^{-3}$ &$4.76\cdot10^{-3}$ &$3.90\cdot10^{-3}$ \\
\end{tabular}
 \end{ruledtabular}
\end{table*}

Inspection of the energies in tables~\ref{tab:energiesa1} and
\ref{tab:energiesa05} and figure~\ref{fig:gausEofa} reveal a clear pattern as all hamiltonians
exhibit rather similar spectra, except for the ones with geometric
potential terms. We emphasize that the tables give absolute values for the
ground state energies, and for better comparison the excitation
energies are given for the excited states.

Let us first consider the hamiltonians without the additional
geometric potential.  Both absolute values and excitation energies for
$H_{EM_1}$ and $H_{EM_2}$ deviate from each other by at most 50\% and the JWKB
approximation always assume intermediate values.  All three energy
sets approach each other with increasing excitation energy in
accordance with an approach towards the validity of classical physics.

The simplest results are from the analytic JWKB approximation which
precisely is the $n^2$ spectrum from a one-dimensional square well.
We know from Eq.~(\ref{wkbenergy}) that an overall energy scale is the
square of the inverse length of the wire.  This is seen as a constant
ratio of $1.160$ for JWKB energies of the same state.  Applying this
scaling from the JWKB approximation on the $H_{EM_1}$ and $H_{EM_2}$ spectra show
the expected decreasing deviations for increasing excitation energies.
These similarities already at the lowest energies are due to the
relatively small coordinate variation of the effective mass.

Including the geometric potential leads to substantially different
spectra.  This potential is overall attractive with constant curvature
outside the bulge region in the center.  The curvature and the
attraction is smaller in the central region.  This has strong
implications on the resulting spectra where the lowest four states
considered here all turn out to have negative energies. In figure~\ref{fig:gausEofa} one sees how for larger $a$ the states turns into two sets of degenerate states. This is because of the barrier in the geometric potential that shows up for larger $a$.

Both JWKB and full solutions with the geometric potential have doubly
degenerate ground states.  This is due to the separation in two
regions through the less attractive central peak of the geometric
potential.  There is room for bound states in each region, and no
distinction between odd and even parity states.  This is highlighted
by the classically forbidden central region which is crucial for the
simple JWKB solution.  A better approximation would allow some
tunneling into this barrier region and the degeneracies would be
lifted as indicated by the energies from the full solutions.

The influence of the geometric potential is much larger than the
variation between the $H_{EM_1}$ and $H_{EM_2}$ spectra.  This applies for both
absolute and relative energies and independent of the bulge parameter
$a$.  In the limit of vanishing $a$ the curvature approach a
coordinate independent constant and the geometric potential becomes
consequently constant as well.  This amounts to a shift of all energies
without any structural changes.  The excitation spectra would thus
approach each other.

\subsection{Eigenfunctions}

\begin{figure}
\centering
\begingroup
  \makeatletter
  \providecommand\color[2][]{%
    \GenericError{(gnuplot) \space\space\space\@spaces}{%
      Package color not loaded in conjunction with
      terminal option `colourtext'%
    }{See the gnuplot documentation for explanation.%
    }{Either use 'blacktext' in gnuplot or load the package
      color.sty in LaTeX.}%
    \renewcommand\color[2][]{}%
  }%
  \providecommand\includegraphics[2][]{%
    \GenericError{(gnuplot) \space\space\space\@spaces}{%
      Package graphicx or graphics not loaded%
    }{See the gnuplot documentation for explanation.%
    }{The gnuplot epslatex terminal needs graphicx.sty or graphics.sty.}%
    \renewcommand\includegraphics[2][]{}%
  }%
  \providecommand\rotatebox[2]{#2}%
  \@ifundefined{ifGPcolor}{%
    \newif\ifGPcolor
    \GPcolortrue
  }{}%
  \@ifundefined{ifGPblacktext}{%
    \newif\ifGPblacktext
    \GPblacktextfalse
  }{}%
  \let\gplgaddtomacro\g@addto@macro
  \gdef\gplbacktext{}%
  \gdef\gplfronttext{}%
  \makeatother
  \ifGPblacktext
    \def\colorrgb#1{}%
    \def\colorgray#1{}%
  \else
    \ifGPcolor
      \def\colorrgb#1{\color[rgb]{#1}}%
      \def\colorgray#1{\color[gray]{#1}}%
      \expandafter\def\csname LTw\endcsname{\color{white}}%
      \expandafter\def\csname LTb\endcsname{\color{black}}%
      \expandafter\def\csname LTa\endcsname{\color{black}}%
      \expandafter\def\csname LT0\endcsname{\color[rgb]{1,0,0}}%
      \expandafter\def\csname LT1\endcsname{\color[rgb]{0,1,0}}%
      \expandafter\def\csname LT2\endcsname{\color[rgb]{0,0,1}}%
      \expandafter\def\csname LT3\endcsname{\color[rgb]{1,0,1}}%
      \expandafter\def\csname LT4\endcsname{\color[rgb]{0,1,1}}%
      \expandafter\def\csname LT5\endcsname{\color[rgb]{1,1,0}}%
      \expandafter\def\csname LT6\endcsname{\color[rgb]{0,0,0}}%
      \expandafter\def\csname LT7\endcsname{\color[rgb]{1,0.3,0}}%
      \expandafter\def\csname LT8\endcsname{\color[rgb]{0.5,0.5,0.5}}%
    \else
      \def\colorrgb#1{\color{black}}%
      \def\colorgray#1{\color[gray]{#1}}%
      \expandafter\def\csname LTw\endcsname{\color{white}}%
      \expandafter\def\csname LTb\endcsname{\color{black}}%
      \expandafter\def\csname LTa\endcsname{\color{black}}%
      \expandafter\def\csname LT0\endcsname{\color{black}}%
      \expandafter\def\csname LT1\endcsname{\color{black}}%
      \expandafter\def\csname LT2\endcsname{\color{black}}%
      \expandafter\def\csname LT3\endcsname{\color{black}}%
      \expandafter\def\csname LT4\endcsname{\color{black}}%
      \expandafter\def\csname LT5\endcsname{\color{black}}%
      \expandafter\def\csname LT6\endcsname{\color{black}}%
      \expandafter\def\csname LT7\endcsname{\color{black}}%
      \expandafter\def\csname LT8\endcsname{\color{black}}%
    \fi
  \fi
  \setlength{\unitlength}{0.0500bp}%
  \begin{picture}(4676.00,3400.00)%
    \gplgaddtomacro\gplbacktext{%
      \csname LTb\endcsname%
      \put(1078,704){\makebox(0,0)[r]{\strut{} 0}}%
      \put(1078,1109){\makebox(0,0)[r]{\strut{} 0.05}}%
      \put(1078,1514){\makebox(0,0)[r]{\strut{} 0.10}}%
      \put(1078,1920){\makebox(0,0)[r]{\strut{} 0.15}}%
      \put(1078,2325){\makebox(0,0)[r]{\strut{} 0.20}}%
      \put(1078,2730){\makebox(0,0)[r]{\strut{} 0.25}}%
      \put(1078,3135){\makebox(0,0)[r]{\strut{} 0.30}}%
      \put(1210,484){\makebox(0,0){\strut{}-40}}%
      \put(1594,484){\makebox(0,0){\strut{}-30}}%
      \put(1977,484){\makebox(0,0){\strut{}-20}}%
      \put(2361,484){\makebox(0,0){\strut{}-10}}%
      \put(2745,484){\makebox(0,0){\strut{} 0}}%
      \put(3128,484){\makebox(0,0){\strut{} 10}}%
      \put(3512,484){\makebox(0,0){\strut{} 20}}%
      \put(3895,484){\makebox(0,0){\strut{} 30}}%
      \put(4279,484){\makebox(0,0){\strut{} 40}}%
      \put(176,1919){\rotatebox{-270}{\makebox(0,0){\strut{}$\Psi(\phi)$}}}%
      \put(2744,154){\makebox(0,0){\strut{}$\phi$}}%
    }%
    \gplgaddtomacro\gplfronttext{%
      \csname LTb\endcsname%
      \put(1870,2962){\makebox(0,0)[r]{\strut{}$H_{geo}$}}%
      \csname LTb\endcsname%
      \put(1870,2742){\makebox(0,0)[r]{\strut{}$H_{EM_2}$}}%
      \csname LTb\endcsname%
      \put(3253,2962){\makebox(0,0)[r]{\strut{}$H_{EM_1}$}}%
      \csname LTb\endcsname%
      \put(3253,2742){\makebox(0,0)[r]{\strut{}JWKB}}%
    }%
    \gplgaddtomacro\gplbacktext{%
      \csname LTb\endcsname%
      \put(1078,704){\makebox(0,0)[r]{\strut{} 0}}%
      \put(1078,1109){\makebox(0,0)[r]{\strut{} 0.05}}%
      \put(1078,1514){\makebox(0,0)[r]{\strut{} 0.10}}%
      \put(1078,1920){\makebox(0,0)[r]{\strut{} 0.15}}%
      \put(1078,2325){\makebox(0,0)[r]{\strut{} 0.20}}%
      \put(1078,2730){\makebox(0,0)[r]{\strut{} 0.25}}%
      \put(1078,3135){\makebox(0,0)[r]{\strut{} 0.30}}%
      \put(1210,484){\makebox(0,0){\strut{}-40}}%
      \put(1594,484){\makebox(0,0){\strut{}-30}}%
      \put(1977,484){\makebox(0,0){\strut{}-20}}%
      \put(2361,484){\makebox(0,0){\strut{}-10}}%
      \put(2745,484){\makebox(0,0){\strut{} 0}}%
      \put(3128,484){\makebox(0,0){\strut{} 10}}%
      \put(3512,484){\makebox(0,0){\strut{} 20}}%
      \put(3895,484){\makebox(0,0){\strut{} 30}}%
      \put(4279,484){\makebox(0,0){\strut{} 40}}%
      \put(176,1919){\rotatebox{-270}{\makebox(0,0){\strut{}$\Psi(\phi)$}}}%
      \put(2744,154){\makebox(0,0){\strut{}$\phi$}}%
    }%
    \gplgaddtomacro\gplfronttext{%
      \csname LTb\endcsname%
      \put(2398,2539){\makebox(0,0)[r]{\strut{}JWKB $+V_{geo}$}}%
    }%
    \gplbacktext
    \put(0,0){\includegraphics{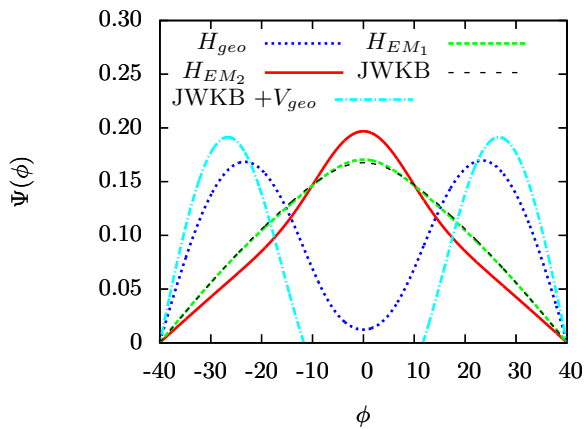}}%
    \gplfronttext
  \end{picture}%
\endgroup
\caption{Top: Ground state wave functions for the different choices
  of Hamiltonian, Eq.(\ref{eq:H2}) (red), Eq.(\ref{eq:pmp}) (green)
  Eq.(\ref{wkbwave}) (black), Eq.(\ref{eq:fullHgeo}) (blue). The
  constants in the parametrization are chosen to be $a=1$ and
  $\phi_0=4\pi$.  Bottom: The effective potentials for $H_{EM_2}$ (red)
  and $H_{geo}$ (blue) for the curve in Eq.(\ref{eq:gausdef}) with
  $a=1$ and $\phi_0=4\pi$
 \label{fig:gausa1b4pin1} }
\end{figure}

The complete picture requires spectra supplemented by corresponding
eigenfunctions.  The ground state wavefunctions for the parameter
choice $a=1$ and $\phi_0=4\pi$ are shown in
fig.~\ref{fig:gausa1b4pin1} for the different hamiltonians.  All the
solutions must vanish at the end of the wire, since this is the
imposed boundary conditions.  

Let us again first consider the cases without geometric potential.
The ground state solutions are all symmetric around the center where
the radius is largest and the largest probabilities occur as peaks.
The solutions to $H_{EM_1}$ in Eq.~(\ref{eq:pmp}) and the JWKB result from
Eq~(\ref{wkbwav}) are almost indistinguishable.  The probability
distributions are rather broad and extending far beyond the bulge in
the central region of the curve.  The solution for $H_{EM_2}$ in
Eq.~(\ref{eq:H2}) is rather similar although distinguishable
with a central peak, slightly higher and correspondingly narrower.

The difference between the $H_{EM_1}$ and $H_{EM_2}$ solutions arises from the
potential in Eq.~(\ref{eq:H1-H2}),
$\frac{\partial^2}{\partial\phi^2}\left(\frac{1}{m(\phi)}\right)$,
which is plotted in fig.~\ref{fig:gausmas}.  This potential is zero at
both ends, becomes repulsive when moving towards the center from
either side of the potential, and finally it turns attractive in the
center.  The wavefunction exhibits a sharp upwards turn when the
attraction is felt with the result of a larger maximum than for $H_{EM_1}$. 

Inclusion of the geometric potential changes qualitatively all the
derived solutions.  The full numerical solution to $H_{geo}$ is still
symmetric but with an almost vanishing minimum in the center and two
prominent maxima on either side.  This behavior is a direct reflection
of the properties of the geometric potential shown in
fig.~\ref{fig:gausmas} and contained in $H_{geo}$.  The JWKB solution
now has two separated classically allowed regions, of course provided
the energies are below the barrier in the center.  The simplest JWKB
solutions are then zero in the central forbidden region as seen in
fig.~\ref{fig:gausa1b4pin1}.  The deviation from the full solution is
therefore very striking but uninteresting since the probabilities at
the same time are very small.  An improved JWKB solution could be
designed by use of an exponentially decreasing wave function to
describe tunneling into the barrier.

\begin{figure}
\centering
\begingroup
  \makeatletter
  \providecommand\color[2][]{%
    \GenericError{(gnuplot) \space\space\space\@spaces}{%
      Package color not loaded in conjunction with
      terminal option `colourtext'%
    }{See the gnuplot documentation for explanation.%
    }{Either use 'blacktext' in gnuplot or load the package
      color.sty in LaTeX.}%
    \renewcommand\color[2][]{}%
  }%
  \providecommand\includegraphics[2][]{%
    \GenericError{(gnuplot) \space\space\space\@spaces}{%
      Package graphicx or graphics not loaded%
    }{See the gnuplot documentation for explanation.%
    }{The gnuplot epslatex terminal needs graphicx.sty or graphics.sty.}%
    \renewcommand\includegraphics[2][]{}%
  }%
  \providecommand\rotatebox[2]{#2}%
  \@ifundefined{ifGPcolor}{%
    \newif\ifGPcolor
    \GPcolortrue
  }{}%
  \@ifundefined{ifGPblacktext}{%
    \newif\ifGPblacktext
    \GPblacktextfalse
  }{}%
  \let\gplgaddtomacro\g@addto@macro
  \gdef\gplbacktext{}%
  \gdef\gplfronttext{}%
  \makeatother
  \ifGPblacktext
    \def\colorrgb#1{}%
    \def\colorgray#1{}%
  \else
    \ifGPcolor
      \def\colorrgb#1{\color[rgb]{#1}}%
      \def\colorgray#1{\color[gray]{#1}}%
      \expandafter\def\csname LTw\endcsname{\color{white}}%
      \expandafter\def\csname LTb\endcsname{\color{black}}%
      \expandafter\def\csname LTa\endcsname{\color{black}}%
      \expandafter\def\csname LT0\endcsname{\color[rgb]{1,0,0}}%
      \expandafter\def\csname LT1\endcsname{\color[rgb]{0,1,0}}%
      \expandafter\def\csname LT2\endcsname{\color[rgb]{0,0,1}}%
      \expandafter\def\csname LT3\endcsname{\color[rgb]{1,0,1}}%
      \expandafter\def\csname LT4\endcsname{\color[rgb]{0,1,1}}%
      \expandafter\def\csname LT5\endcsname{\color[rgb]{1,1,0}}%
      \expandafter\def\csname LT6\endcsname{\color[rgb]{0,0,0}}%
      \expandafter\def\csname LT7\endcsname{\color[rgb]{1,0.3,0}}%
      \expandafter\def\csname LT8\endcsname{\color[rgb]{0.5,0.5,0.5}}%
    \else
      \def\colorrgb#1{\color{black}}%
      \def\colorgray#1{\color[gray]{#1}}%
      \expandafter\def\csname LTw\endcsname{\color{white}}%
      \expandafter\def\csname LTb\endcsname{\color{black}}%
      \expandafter\def\csname LTa\endcsname{\color{black}}%
      \expandafter\def\csname LT0\endcsname{\color{black}}%
      \expandafter\def\csname LT1\endcsname{\color{black}}%
      \expandafter\def\csname LT2\endcsname{\color{black}}%
      \expandafter\def\csname LT3\endcsname{\color{black}}%
      \expandafter\def\csname LT4\endcsname{\color{black}}%
      \expandafter\def\csname LT5\endcsname{\color{black}}%
      \expandafter\def\csname LT6\endcsname{\color{black}}%
      \expandafter\def\csname LT7\endcsname{\color{black}}%
      \expandafter\def\csname LT8\endcsname{\color{black}}%
    \fi
  \fi
  \setlength{\unitlength}{0.0500bp}%
  \begin{picture}(4676.00,3400.00)%
    \gplgaddtomacro\gplbacktext{%
      \csname LTb\endcsname%
      \put(946,704){\makebox(0,0)[r]{\strut{}-0.2}}%
      \put(946,1146){\makebox(0,0)[r]{\strut{}-0.1}}%
      \put(946,1588){\makebox(0,0)[r]{\strut{} 0}}%
      \put(946,2030){\makebox(0,0)[r]{\strut{} 0.1}}%
      \put(946,2472){\makebox(0,0)[r]{\strut{} 0.2}}%
      \put(946,2914){\makebox(0,0)[r]{\strut{} 0.3}}%
      \put(1078,484){\makebox(0,0){\strut{}-40}}%
      \put(1478,484){\makebox(0,0){\strut{}-30}}%
      \put(1878,484){\makebox(0,0){\strut{}-20}}%
      \put(2278,484){\makebox(0,0){\strut{}-10}}%
      \put(2679,484){\makebox(0,0){\strut{} 0}}%
      \put(3079,484){\makebox(0,0){\strut{} 10}}%
      \put(3479,484){\makebox(0,0){\strut{} 20}}%
      \put(3879,484){\makebox(0,0){\strut{} 30}}%
      \put(4279,484){\makebox(0,0){\strut{} 40}}%
      \put(176,1919){\rotatebox{-270}{\makebox(0,0){\strut{}$\psi(\phi)$}}}%
      \put(2678,154){\makebox(0,0){\strut{}$\phi$}}%
    }%
    \gplgaddtomacro\gplfronttext{%
      \csname LTb\endcsname%
      \put(1738,2962){\makebox(0,0)[r]{\strut{}$H_{geo}$}}%
      \csname LTb\endcsname%
      \put(1738,2742){\makebox(0,0)[r]{\strut{}$H_{EM_2}$}}%
      \csname LTb\endcsname%
      \put(3121,2962){\makebox(0,0)[r]{\strut{}$H_{EM_1}$}}%
      \csname LTb\endcsname%
      \put(3121,2742){\makebox(0,0)[r]{\strut{}JWKB}}%
    }%
    \gplgaddtomacro\gplbacktext{%
      \csname LTb\endcsname%
      \put(946,704){\makebox(0,0)[r]{\strut{}-0.2}}%
      \put(946,1146){\makebox(0,0)[r]{\strut{}-0.1}}%
      \put(946,1588){\makebox(0,0)[r]{\strut{} 0}}%
      \put(946,2030){\makebox(0,0)[r]{\strut{} 0.1}}%
      \put(946,2472){\makebox(0,0)[r]{\strut{} 0.2}}%
      \put(946,2914){\makebox(0,0)[r]{\strut{} 0.3}}%
      \put(1078,484){\makebox(0,0){\strut{}-40}}%
      \put(1478,484){\makebox(0,0){\strut{}-30}}%
      \put(1878,484){\makebox(0,0){\strut{}-20}}%
      \put(2278,484){\makebox(0,0){\strut{}-10}}%
      \put(2679,484){\makebox(0,0){\strut{} 0}}%
      \put(3079,484){\makebox(0,0){\strut{} 10}}%
      \put(3479,484){\makebox(0,0){\strut{} 20}}%
      \put(3879,484){\makebox(0,0){\strut{} 30}}%
      \put(4279,484){\makebox(0,0){\strut{} 40}}%
      \put(176,1919){\rotatebox{-270}{\makebox(0,0){\strut{}$\psi(\phi)$}}}%
      \put(2678,154){\makebox(0,0){\strut{}$\phi$}}%
    }%
    \gplgaddtomacro\gplfronttext{%
      \csname LTb\endcsname%
      \put(2266,2539){\makebox(0,0)[r]{\strut{}JWKB $+V_{geo}$}}%
    }%
    \gplbacktext
    \put(0,0){\includegraphics{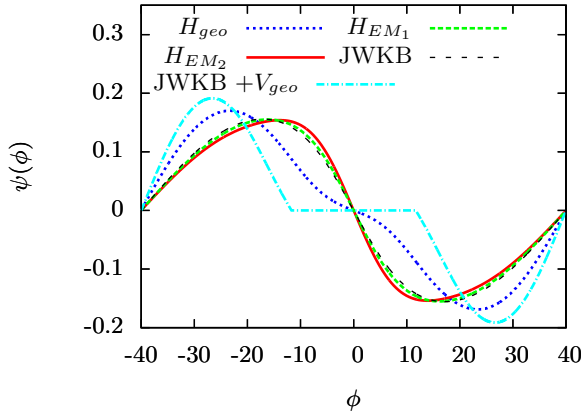}}%
    \gplfronttext
  \end{picture}%
\endgroup
\caption{First excited states for the different choices of
  Hamiltonian, Eq.(\ref{eq:H2}) (red), Eq.(\ref{eq:pmp}) (green),
  Eq.(\ref{wkbwave}) (black). The constants in the
  parametrization are chosen to be $a=1$ and $\phi_0=4\pi$.
 \label{fig:gausa1b4pin2}}
\end{figure}

The first excited states are shown in fig.~\ref{fig:gausa1b4pin2}
for the different hamiltonians.  The boundary conditions of zero at
the end points of the wire are maintained in all cases.  Now a node
appears in the center, and all wave functions are of odd parity.  The
solutions without geometric potential are almost indistinguishable for
the $H_{EM_1}$, $H_{EM_2}$, solutions and the related JWKB approximation.  They all
resemble the sine wave functions for the first excited state of a
particle in a square well potential.  Now the central attraction for
$H_{EM_2}$ is much less effective due to the required node for $\phi = 0$.

Again including the geometric potential changes the solutions
substantially, although much less than for the ground state.  The odd
parity characteristics with a central node is maintained but now with
a small oscillatory modulation by tunneling into the barrier.  The two
regions of large probability are pushed further away from the center
by the potential than the $H_{EM_1}$ and $H_{EM_2}$ solutions. The slope of the
wave functions is also not as steep across the central region. 

The corresponding simplest JWKB solution is trying to mimic this
behavior in the classically allowed regions.  The central forbidden
region has a constant probability of zero.  The only difference between
first excited and ground state is that the ground state is even, and
the first excited state is odd.  The absolute values corresponding to
the probability distributions would be identical.

\begin{figure}
\centering
\begingroup
  \makeatletter
  \providecommand\color[2][]{%
    \GenericError{(gnuplot) \space\space\space\@spaces}{%
      Package color not loaded in conjunction with
      terminal option `colourtext'%
    }{See the gnuplot documentation for explanation.%
    }{Either use 'blacktext' in gnuplot or load the package
      color.sty in LaTeX.}%
    \renewcommand\color[2][]{}%
  }%
  \providecommand\includegraphics[2][]{%
    \GenericError{(gnuplot) \space\space\space\@spaces}{%
      Package graphicx or graphics not loaded%
    }{See the gnuplot documentation for explanation.%
    }{The gnuplot epslatex terminal needs graphicx.sty or graphics.sty.}%
    \renewcommand\includegraphics[2][]{}%
  }%
  \providecommand\rotatebox[2]{#2}%
  \@ifundefined{ifGPcolor}{%
    \newif\ifGPcolor
    \GPcolortrue
  }{}%
  \@ifundefined{ifGPblacktext}{%
    \newif\ifGPblacktext
    \GPblacktextfalse
  }{}%
  \let\gplgaddtomacro\g@addto@macro
  \gdef\gplbacktext{}%
  \gdef\gplfronttext{}%
  \makeatother
  \ifGPblacktext
    \def\colorrgb#1{}%
    \def\colorgray#1{}%
  \else
    \ifGPcolor
      \def\colorrgb#1{\color[rgb]{#1}}%
      \def\colorgray#1{\color[gray]{#1}}%
      \expandafter\def\csname LTw\endcsname{\color{white}}%
      \expandafter\def\csname LTb\endcsname{\color{black}}%
      \expandafter\def\csname LTa\endcsname{\color{black}}%
      \expandafter\def\csname LT0\endcsname{\color[rgb]{1,0,0}}%
      \expandafter\def\csname LT1\endcsname{\color[rgb]{0,1,0}}%
      \expandafter\def\csname LT2\endcsname{\color[rgb]{0,0,1}}%
      \expandafter\def\csname LT3\endcsname{\color[rgb]{1,0,1}}%
      \expandafter\def\csname LT4\endcsname{\color[rgb]{0,1,1}}%
      \expandafter\def\csname LT5\endcsname{\color[rgb]{1,1,0}}%
      \expandafter\def\csname LT6\endcsname{\color[rgb]{0,0,0}}%
      \expandafter\def\csname LT7\endcsname{\color[rgb]{1,0.3,0}}%
      \expandafter\def\csname LT8\endcsname{\color[rgb]{0.5,0.5,0.5}}%
    \else
      \def\colorrgb#1{\color{black}}%
      \def\colorgray#1{\color[gray]{#1}}%
      \expandafter\def\csname LTw\endcsname{\color{white}}%
      \expandafter\def\csname LTb\endcsname{\color{black}}%
      \expandafter\def\csname LTa\endcsname{\color{black}}%
      \expandafter\def\csname LT0\endcsname{\color{black}}%
      \expandafter\def\csname LT1\endcsname{\color{black}}%
      \expandafter\def\csname LT2\endcsname{\color{black}}%
      \expandafter\def\csname LT3\endcsname{\color{black}}%
      \expandafter\def\csname LT4\endcsname{\color{black}}%
      \expandafter\def\csname LT5\endcsname{\color{black}}%
      \expandafter\def\csname LT6\endcsname{\color{black}}%
      \expandafter\def\csname LT7\endcsname{\color{black}}%
      \expandafter\def\csname LT8\endcsname{\color{black}}%
    \fi
  \fi
  \setlength{\unitlength}{0.0500bp}%
  \begin{picture}(4676.00,3400.00)%
    \gplgaddtomacro\gplbacktext{%
      \csname LTb\endcsname%
      \put(1078,704){\makebox(0,0)[r]{\strut{} 0}}%
      \put(1078,1109){\makebox(0,0)[r]{\strut{} 0.05}}%
      \put(1078,1514){\makebox(0,0)[r]{\strut{} 0.10}}%
      \put(1078,1920){\makebox(0,0)[r]{\strut{} 0.15}}%
      \put(1078,2325){\makebox(0,0)[r]{\strut{} 0.20}}%
      \put(1078,2730){\makebox(0,0)[r]{\strut{} 0.25}}%
      \put(1078,3135){\makebox(0,0)[r]{\strut{} 0.30}}%
      \put(1210,484){\makebox(0,0){\strut{}-40}}%
      \put(1594,484){\makebox(0,0){\strut{}-30}}%
      \put(1977,484){\makebox(0,0){\strut{}-20}}%
      \put(2361,484){\makebox(0,0){\strut{}-10}}%
      \put(2745,484){\makebox(0,0){\strut{} 0}}%
      \put(3128,484){\makebox(0,0){\strut{} 10}}%
      \put(3512,484){\makebox(0,0){\strut{} 20}}%
      \put(3895,484){\makebox(0,0){\strut{} 30}}%
      \put(4279,484){\makebox(0,0){\strut{} 40}}%
      \put(176,1919){\rotatebox{-270}{\makebox(0,0){\strut{}$\Psi(\phi)$}}}%
      \put(2744,154){\makebox(0,0){\strut{}$\phi$}}%
    }%
    \gplgaddtomacro\gplfronttext{%
      \csname LTb\endcsname%
      \put(1870,2962){\makebox(0,0)[r]{\strut{}$H_{geo}$}}%
      \csname LTb\endcsname%
      \put(1870,2742){\makebox(0,0)[r]{\strut{}$H_{EM_2}$}}%
      \csname LTb\endcsname%
      \put(3253,2962){\makebox(0,0)[r]{\strut{}$H_{EM_1}$}}%
      \csname LTb\endcsname%
      \put(3253,2742){\makebox(0,0)[r]{\strut{}JWKB}}%
    }%
    \gplgaddtomacro\gplbacktext{%
      \csname LTb\endcsname%
      \put(1078,704){\makebox(0,0)[r]{\strut{} 0}}%
      \put(1078,1109){\makebox(0,0)[r]{\strut{} 0.05}}%
      \put(1078,1514){\makebox(0,0)[r]{\strut{} 0.10}}%
      \put(1078,1920){\makebox(0,0)[r]{\strut{} 0.15}}%
      \put(1078,2325){\makebox(0,0)[r]{\strut{} 0.20}}%
      \put(1078,2730){\makebox(0,0)[r]{\strut{} 0.25}}%
      \put(1078,3135){\makebox(0,0)[r]{\strut{} 0.30}}%
      \put(1210,484){\makebox(0,0){\strut{}-40}}%
      \put(1594,484){\makebox(0,0){\strut{}-30}}%
      \put(1977,484){\makebox(0,0){\strut{}-20}}%
      \put(2361,484){\makebox(0,0){\strut{}-10}}%
      \put(2745,484){\makebox(0,0){\strut{} 0}}%
      \put(3128,484){\makebox(0,0){\strut{} 10}}%
      \put(3512,484){\makebox(0,0){\strut{} 20}}%
      \put(3895,484){\makebox(0,0){\strut{} 30}}%
      \put(4279,484){\makebox(0,0){\strut{} 40}}%
      \put(176,1919){\rotatebox{-270}{\makebox(0,0){\strut{}$\Psi(\phi)$}}}%
      \put(2744,154){\makebox(0,0){\strut{}$\phi$}}%
    }%
    \gplgaddtomacro\gplfronttext{%
      \csname LTb\endcsname%
      \put(2398,2539){\makebox(0,0)[r]{\strut{}JWKB $+V_{geo}$}}%
    }%
    \gplbacktext
    \put(0,0){\includegraphics{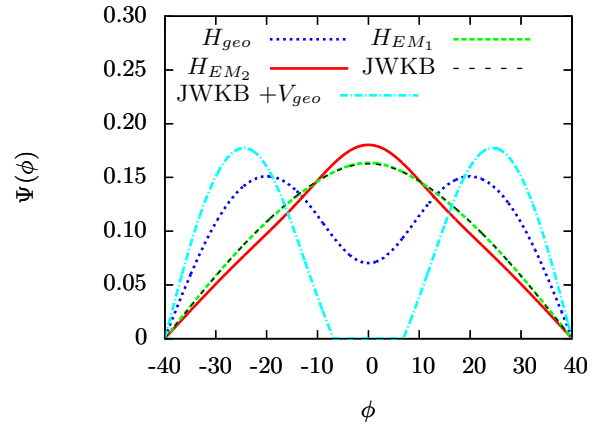}}%
    \gplfronttext
  \end{picture}%
\endgroup
\caption{Top: Ground states for the different choices of Hamiltonian,
  Eq.(\ref{eq:H2}) (red), Eq.(\ref{eq:pmp}) (green),
  Eq.(\ref{wkbwave}) (black). The constants in the parametrization are
  chosen to be $a=0.5$ and $\phi_0=4\pi$.  Bottom: The effective
  potentials for $H_{EM_2}$(red) and $H_{geo}$(blue) for the curve in
  Eq.(\ref{eq:gausdef}) with $a=0.5$ and $\phi_0=4\pi$
 \label{fig:gausa05b4pin1} }
\end{figure}

The dependence of the eigenfunctions on the parameters of the gaussian
central bump is intuitively clear with the detailed knowledge we
accumulated from the investigated set.  The effects from varying the
two parameters, $a$ and $\phi_0$, seems to be very different.  The
value of $aR$ is directly the radial extension of the bump beyond
initial helix radius. It is therefore clear that increasing $a$ from
zero must increase the change of the solutions from the $a=0$
solutions where both mass and curvature are constants.  However, the
qualitative behavior of even and odd parity is maintained, and with
the related maxima or nodes at the center.

The $\phi_0$ variation appears to be very different but the effects
are actually rather similar.  Large values imply a slow variation of
the radius of the helix, and as such small influence on the wave
functions beyond a possible scaling from a different average radius.
Small $\phi_0$ values imply fast variation over few windings. Both
mass and curvature would then vary much faster as well, and the
different quantizations would be very different.

Thus, large $a$ and small $\phi_0$ lead to large variation in mass and
curvature and consequently the different quantization prescriptions
would deviate more and more from each other. This would be
particularly prominent in comparison with use of the geometric
potential.  It is worth to emphasize that it is not obvious which
quantization procedure is most correct for these one-dimensional cases. 

On one hand the simple JWKB approximation provides a very accurate
match with the $H_{EM_1}$ solutions. However, this assumes that no
potential is necessary to confine the particle to the one-dimensional
wire.  On the other hand, some geometric potential combined with an
appropriate kinetic energy operator would directly deliver the
hamiltonian.  Unfortunately, this assumes a computational scheme to
obtain a reliable potential, and the lowest order curvature dependent
potential is not accurate for helix like periodic structures with
strongly varying effective mass.

\section{Stretched helix \label{sec:stretch}}

In contrast to the previous section we shall here investigate
asymmetric helix deformations. We design here two stretching
parameterizations originating from very different assumptions. We first
describe these parameterizations, and in the following subsections we
present results for spectra and eigenfunctions for the different
quantization descriptions.

\subsection{Parameterizations}

\begin{figure}
\centering
\includegraphics[width=\columnwidth]{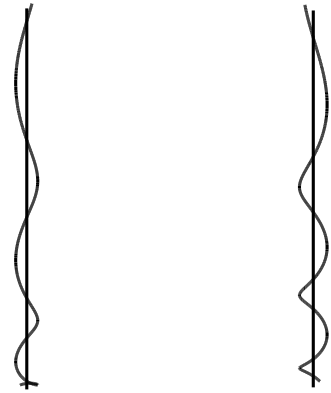}
\caption{Schematic drawings of two deformed helices with stretched and
  squeezed parametrizations Eqs.~(\ref{eq:massstre}) and
  (\ref{eq:inversedef}).  The stretching and squeezing parameters are
  chosen to be $a= 0.1$, and $c_0 = c_1 = 0.01$.
 \label{fig:helix2}}
\end{figure}

We want to study the non-trivial cases where both mass and curvature
are monotonously varying with the coordinate.  Instead of symmetry we
choose an increasing stretching along the symmetry axis of the helix,
that is
\begin{eqnarray}
f_x(\phi)= f_y(\phi) \;\; ,\;\; f_z(\phi)= a\phi^2 \;,
\label{eq:phisquared}
\end{eqnarray}
where $a$ is a constant. This curve is circular in the $x-y$ plane and
the distance between the windings increase with the angle, $\phi$, see
fig.~\ref{fig:helix2}.  The effective mass is simple, that is
\begin{eqnarray}
 m(\phi)=m_0R^2\left(1+4a^2\phi^2\right) \; ,
\label{eq:massstre}
\end{eqnarray}
which allow analytical integration of the square root in
Eq.~(\ref{eq:wkbquant}),  and therefore a fully analytic JWKB-solution.
Only first and second derivatives are finite and expansion in higher
order derivatives are more likely to converge than for a periodic
structure.  

Explicitly we get the bound state wave function given by
Eq.~(\ref{wkbwave}) with
\begin{eqnarray} \label{wkbwav}
& &  L(\phi) = R\left(\frac{1}{4a}
 \ln{\left[\frac{\phi+\sqrt{\phi^2+\frac{1}{4a^2}}}
 {\phi_{min}+\sqrt{\phi_{min}^2+\frac{1}{4a^2}}}\right]}\right. \nonumber\\ 
& & \left. + a\phi\sqrt{\phi^2+\frac{1}{4a^2}}-
  a \phi_{min}\sqrt{\phi_{min}^2+\frac{1}{4a^2}}\right) \;,
\label{wkblength}
\end{eqnarray}  
where $L(\phi_{max})$ is the length of the wire.

The two hamiltonians $H_{EM_1}$ and $H_{EM_2}$ differ by the second derivative
of the inverse mass, see Eq.~(\ref{eq:H1-H2}).  We can find a
parametrization where this difference vanishes.  The assumption of
identical circles in the $x-y$ plane, $f_x=f_y=1$, gives an effective
mass from Eq.~(\ref{eq:mdef}), $m(\phi) = m_0R^2 (1+ (f'_z)^2) $.  If we
therefore assume that $H_{EM_1}=H_{EM_2}$ then $1/(1+ (f'_z)^2)$ should be a first
order polynomium i $\phi$, or equivalently
$m(\phi)=\frac{1}{c_0+c_1\phi}$, where $c_0$ and $c_1$ are constants.

Integrating $f'_z(\phi)$ to find $f_z(\phi)$, we get the
parametrization:
\begin{eqnarray}
 f_x(\phi)&=&  f_y(\phi) = 1 \nonumber \\ \nonumber
 f_z(\phi)&=&-\frac{\arctan{\left[\frac{-2c_0-2c_1\phi+1}{2\sqrt{-\left(c_0+c_1\phi-1\right)\left(c_0+c_1\phi\right)}}\right]}}{2c_1}\\
&-&\frac{2R\sqrt{-\left(c_0+c_1\phi-1\right)\left(c_0+c_1\phi\right)}}{2c_1} \;.
\label{eq:inversedef}
\end{eqnarray}
The curves in Eqs.~(\ref{eq:massstre}) and (\ref{eq:inversedef}) are
different types of monotonous deformations in the z-direction.  The
resulting deformed helices are both shown in fig.~\ref{fig:helix2}.

\begin{figure}
\centering
\begingroup
  \makeatletter
  \providecommand\color[2][]{%
    \GenericError{(gnuplot) \space\space\space\@spaces}{%
      Package color not loaded in conjunction with
      terminal option `colourtext'%
    }{See the gnuplot documentation for explanation.%
    }{Either use 'blacktext' in gnuplot or load the package
      color.sty in LaTeX.}%
    \renewcommand\color[2][]{}%
  }%
  \providecommand\includegraphics[2][]{%
    \GenericError{(gnuplot) \space\space\space\@spaces}{%
      Package graphicx or graphics not loaded%
    }{See the gnuplot documentation for explanation.%
    }{The gnuplot epslatex terminal needs graphicx.sty or graphics.sty.}%
    \renewcommand\includegraphics[2][]{}%
  }%
  \providecommand\rotatebox[2]{#2}%
  \@ifundefined{ifGPcolor}{%
    \newif\ifGPcolor
    \GPcolortrue
  }{}%
  \@ifundefined{ifGPblacktext}{%
    \newif\ifGPblacktext
    \GPblacktextfalse
  }{}%
  \let\gplgaddtomacro\g@addto@macro
  \gdef\gplbacktext{}%
  \gdef\gplfronttext{}%
  \makeatother
  \ifGPblacktext
    \def\colorrgb#1{}%
    \def\colorgray#1{}%
  \else
    \ifGPcolor
      \def\colorrgb#1{\color[rgb]{#1}}%
      \def\colorgray#1{\color[gray]{#1}}%
      \expandafter\def\csname LTw\endcsname{\color{white}}%
      \expandafter\def\csname LTb\endcsname{\color{black}}%
      \expandafter\def\csname LTa\endcsname{\color{black}}%
      \expandafter\def\csname LT0\endcsname{\color[rgb]{1,0,0}}%
      \expandafter\def\csname LT1\endcsname{\color[rgb]{0,1,0}}%
      \expandafter\def\csname LT2\endcsname{\color[rgb]{0,0,1}}%
      \expandafter\def\csname LT3\endcsname{\color[rgb]{1,0,1}}%
      \expandafter\def\csname LT4\endcsname{\color[rgb]{0,1,1}}%
      \expandafter\def\csname LT5\endcsname{\color[rgb]{1,1,0}}%
      \expandafter\def\csname LT6\endcsname{\color[rgb]{0,0,0}}%
      \expandafter\def\csname LT7\endcsname{\color[rgb]{1,0.3,0}}%
      \expandafter\def\csname LT8\endcsname{\color[rgb]{0.5,0.5,0.5}}%
    \else
      \def\colorrgb#1{\color{black}}%
      \def\colorgray#1{\color[gray]{#1}}%
      \expandafter\def\csname LTw\endcsname{\color{white}}%
      \expandafter\def\csname LTb\endcsname{\color{black}}%
      \expandafter\def\csname LTa\endcsname{\color{black}}%
      \expandafter\def\csname LT0\endcsname{\color{black}}%
      \expandafter\def\csname LT1\endcsname{\color{black}}%
      \expandafter\def\csname LT2\endcsname{\color{black}}%
      \expandafter\def\csname LT3\endcsname{\color{black}}%
      \expandafter\def\csname LT4\endcsname{\color{black}}%
      \expandafter\def\csname LT5\endcsname{\color{black}}%
      \expandafter\def\csname LT6\endcsname{\color{black}}%
      \expandafter\def\csname LT7\endcsname{\color{black}}%
      \expandafter\def\csname LT8\endcsname{\color{black}}%
    \fi
  \fi
  \setlength{\unitlength}{0.0500bp}%
  \begin{picture}(3968.00,3968.00)%
    \gplgaddtomacro\gplbacktext{%
      \csname LTb\endcsname%
      \put(66,1984){\makebox(0,0)[r]{\strut{} 0}}%
      \put(66,2292){\makebox(0,0)[r]{\strut{} 400}}%
      \put(66,2600){\makebox(0,0)[r]{\strut{} 800}}%
      \put(66,2909){\makebox(0,0)[r]{\strut{} 1200}}%
      \put(66,3217){\makebox(0,0)[r]{\strut{} 1600}}%
    }%
    \gplgaddtomacro\gplfronttext{%
      \csname LTb\endcsname%
      \put(1122,3198){\makebox(0,0)[r]{\strut{}$m(x)$}}%
      \csname LTb\endcsname%
      \put(1122,2978){\makebox(0,0)[r]{\strut{}$m^,(x)$}}%
      \csname LTb\endcsname%
      \put(1122,2758){\makebox(0,0)[r]{\strut{}$m^{,,}(x)$}}%
    }%
    \gplgaddtomacro\gplbacktext{%
      \csname LTb\endcsname%
      \put(66,595){\makebox(0,0)[r]{\strut{}-8}}%
      \put(66,780){\makebox(0,0)[r]{\strut{}-6}}%
      \put(66,965){\makebox(0,0)[r]{\strut{}-4}}%
      \put(66,1150){\makebox(0,0)[r]{\strut{}-2}}%
      \put(66,1335){\makebox(0,0)[r]{\strut{} 0}}%
      \put(66,1520){\makebox(0,0)[r]{\strut{} 2}}%
      \put(66,1705){\makebox(0,0)[r]{\strut{} 4}}%
      \put(198,375){\makebox(0,0){\strut{} 0}}%
      \put(1091,375){\makebox(0,0){\strut{} 5}}%
      \put(1983,375){\makebox(0,0){\strut{} 10}}%
      \put(2876,375){\makebox(0,0){\strut{} 15}}%
      \put(3768,375){\makebox(0,0){\strut{} 20}}%
      \put(1983,45){\makebox(0,0){\strut{}$\phi$}}%
    }%
    \gplgaddtomacro\gplfronttext{%
      \csname LTb\endcsname%
      \put(858,1810){\makebox(0,0)[r]{\strut{}$V_{EM_2}$}}%
      \csname LTb\endcsname%
      \put(858,1590){\makebox(0,0)[r]{\strut{}$V_{geo}$}}%
    }%
    \gplgaddtomacro\gplbacktext{%
      \csname LTb\endcsname%
      \put(66,595){\makebox(0,0)[r]{\strut{}-8}}%
      \put(66,780){\makebox(0,0)[r]{\strut{}-6}}%
      \put(66,965){\makebox(0,0)[r]{\strut{}-4}}%
      \put(66,1150){\makebox(0,0)[r]{\strut{}-2}}%
      \put(66,1335){\makebox(0,0)[r]{\strut{} 0}}%
      \put(66,1520){\makebox(0,0)[r]{\strut{} 2}}%
      \put(66,1705){\makebox(0,0)[r]{\strut{} 4}}%
      \put(198,375){\makebox(0,0){\strut{} 0}}%
      \put(1091,375){\makebox(0,0){\strut{} 5}}%
      \put(1983,375){\makebox(0,0){\strut{} 10}}%
      \put(2876,375){\makebox(0,0){\strut{} 15}}%
      \put(3768,375){\makebox(0,0){\strut{} 20}}%
      \put(1983,45){\makebox(0,0){\strut{}$\phi$}}%
    }%
    \gplgaddtomacro\gplfronttext{%
      \csname LTb\endcsname%
      \put(858,768){\makebox(0,0)[r]{\strut{}$V_{EM_{1}}$}}%
    }%
    \gplbacktext
    \put(0,0){\includegraphics{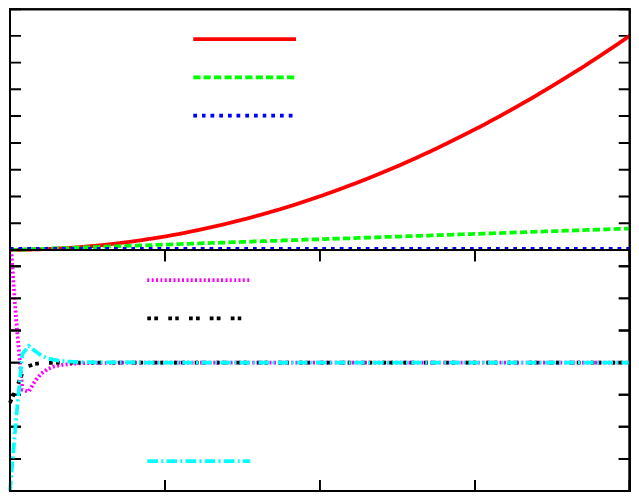}}%
    \gplfronttext
  \end{picture}%
\endgroup
\caption{Mass and corresponding derivatives as function of angle,
  $\phi$, for the parametrization in eq.(\ref{eq:phisquared}).  Top:
  $m(\phi)$ (red), $\cdot m'(\phi)$ (green), $\cdot m''(\phi)$ (blue), all in
  units of $m_0R^2$.  Bottom contains potentials in units of
  $\frac{\hbar^2}{4m_0R^2}$ that enter different quantization
  prescriptions: $-\big(m''-
  \frac{3}{2}\frac{m'^2}{m}\big)/m^2$(magenta), $\big(m''-2\frac{m'^2}{m}\big)/m^2$(cyan), $-\kappa^2/2$(black), see
  Eqs.(\ref{eq:H1red}), (\ref{eq:H1-H2}) and (\ref{eq:geo}).
 \label{fig:helixmasa}}
\end{figure}

\begin{figure}
\centering
\begingroup
  \makeatletter
  \providecommand\color[2][]{%
    \GenericError{(gnuplot) \space\space\space\@spaces}{%
      Package color not loaded in conjunction with
      terminal option `colourtext'%
    }{See the gnuplot documentation for explanation.%
    }{Either use 'blacktext' in gnuplot or load the package
      color.sty in LaTeX.}%
    \renewcommand\color[2][]{}%
  }%
  \providecommand\includegraphics[2][]{%
    \GenericError{(gnuplot) \space\space\space\@spaces}{%
      Package graphicx or graphics not loaded%
    }{See the gnuplot documentation for explanation.%
    }{The gnuplot epslatex terminal needs graphicx.sty or graphics.sty.}%
    \renewcommand\includegraphics[2][]{}%
  }%
  \providecommand\rotatebox[2]{#2}%
  \@ifundefined{ifGPcolor}{%
    \newif\ifGPcolor
    \GPcolortrue
  }{}%
  \@ifundefined{ifGPblacktext}{%
    \newif\ifGPblacktext
    \GPblacktextfalse
  }{}%
  \let\gplgaddtomacro\g@addto@macro
  \gdef\gplbacktext{}%
  \gdef\gplfronttext{}%
  \makeatother
  \ifGPblacktext
    \def\colorrgb#1{}%
    \def\colorgray#1{}%
  \else
    \ifGPcolor
      \def\colorrgb#1{\color[rgb]{#1}}%
      \def\colorgray#1{\color[gray]{#1}}%
      \expandafter\def\csname LTw\endcsname{\color{white}}%
      \expandafter\def\csname LTb\endcsname{\color{black}}%
      \expandafter\def\csname LTa\endcsname{\color{black}}%
      \expandafter\def\csname LT0\endcsname{\color[rgb]{1,0,0}}%
      \expandafter\def\csname LT1\endcsname{\color[rgb]{0,1,0}}%
      \expandafter\def\csname LT2\endcsname{\color[rgb]{0,0,1}}%
      \expandafter\def\csname LT3\endcsname{\color[rgb]{1,0,1}}%
      \expandafter\def\csname LT4\endcsname{\color[rgb]{0,1,1}}%
      \expandafter\def\csname LT5\endcsname{\color[rgb]{1,1,0}}%
      \expandafter\def\csname LT6\endcsname{\color[rgb]{0,0,0}}%
      \expandafter\def\csname LT7\endcsname{\color[rgb]{1,0.3,0}}%
      \expandafter\def\csname LT8\endcsname{\color[rgb]{0.5,0.5,0.5}}%
    \else
      \def\colorrgb#1{\color{black}}%
      \def\colorgray#1{\color[gray]{#1}}%
      \expandafter\def\csname LTw\endcsname{\color{white}}%
      \expandafter\def\csname LTb\endcsname{\color{black}}%
      \expandafter\def\csname LTa\endcsname{\color{black}}%
      \expandafter\def\csname LT0\endcsname{\color{black}}%
      \expandafter\def\csname LT1\endcsname{\color{black}}%
      \expandafter\def\csname LT2\endcsname{\color{black}}%
      \expandafter\def\csname LT3\endcsname{\color{black}}%
      \expandafter\def\csname LT4\endcsname{\color{black}}%
      \expandafter\def\csname LT5\endcsname{\color{black}}%
      \expandafter\def\csname LT6\endcsname{\color{black}}%
      \expandafter\def\csname LT7\endcsname{\color{black}}%
      \expandafter\def\csname LT8\endcsname{\color{black}}%
    \fi
  \fi
  \setlength{\unitlength}{0.0500bp}%
  \begin{picture}(3968.00,3968.00)%
    \gplgaddtomacro\gplbacktext{%
      \csname LTb\endcsname%
      \put(66,2182){\makebox(0,0)[r]{\strut{}-100}}%
      \put(66,2380){\makebox(0,0)[r]{\strut{}-50}}%
      \put(66,2578){\makebox(0,0)[r]{\strut{} 0}}%
      \put(66,2777){\makebox(0,0)[r]{\strut{} 50}}%
      \put(66,2975){\makebox(0,0)[r]{\strut{} 100}}%
      \put(66,3173){\makebox(0,0)[r]{\strut{} 150}}%
      \put(66,3371){\makebox(0,0)[r]{\strut{} 200}}%
    }%
    \gplgaddtomacro\gplfronttext{%
      \csname LTb\endcsname%
      \put(2781,3198){\makebox(0,0)[r]{\strut{}$m(x)$}}%
      \csname LTb\endcsname%
      \put(2781,2978){\makebox(0,0)[r]{\strut{}$m^,(x)$}}%
      \csname LTb\endcsname%
      \put(2781,2758){\makebox(0,0)[r]{\strut{}$m^{,,}(x)$}}%
    }%
    \gplgaddtomacro\gplbacktext{%
      \csname LTb\endcsname%
      \put(66,595){\makebox(0,0)[r]{\strut{}-0.006}}%
      \put(66,809){\makebox(0,0)[r]{\strut{}-0.005}}%
      \put(66,1022){\makebox(0,0)[r]{\strut{}-0.004}}%
      \put(66,1236){\makebox(0,0)[r]{\strut{}-0.003}}%
      \put(66,1449){\makebox(0,0)[r]{\strut{}-0.002}}%
      \put(66,1663){\makebox(0,0)[r]{\strut{}-0.001}}%
      \put(66,1876){\makebox(0,0)[r]{\strut{} 0}}%
      \put(198,375){\makebox(0,0){\strut{} 0}}%
      \put(644,375){\makebox(0,0){\strut{} 5}}%
      \put(1091,375){\makebox(0,0){\strut{} 10}}%
      \put(1537,375){\makebox(0,0){\strut{} 15}}%
      \put(1983,375){\makebox(0,0){\strut{} 20}}%
      \put(2429,375){\makebox(0,0){\strut{} 25}}%
      \put(2876,375){\makebox(0,0){\strut{} 30}}%
      \put(3322,375){\makebox(0,0){\strut{} 35}}%
      \put(3768,375){\makebox(0,0){\strut{} 40}}%
    }%
    \gplgaddtomacro\gplfronttext{%
      \csname LTb\endcsname%
      \put(858,1208){\makebox(0,0)[r]{\strut{}$V_{EM_2}$}}%
      \csname LTb\endcsname%
      \put(858,988){\makebox(0,0)[r]{\strut{}$V_{geo}$}}%
      \csname LTb\endcsname%
      \put(858,768){\makebox(0,0)[r]{\strut{}$V_{EM_{1}}$}}%
    }%
    \gplbacktext
    \put(0,0){\includegraphics{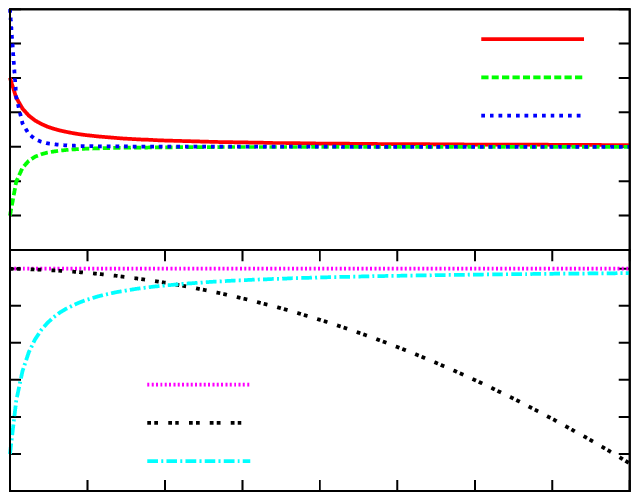}}%
    \gplfronttext
  \end{picture}%
\endgroup
\caption{Mass and corresponding derivatives as function of angle,
  $\phi$, for the parametrization in eq.(\ref{eq:inversedef}). Top:
  $m(\phi)$ (red), $\cdot m'(\phi)$ (green), $\cdot m''(\phi)$ (blue), all in
  units of $m_0R^2$.  Bottom contains potentials in units of
  $\frac{\hbar^2}{4m_0R^2}$ that enter different quantization
  prescriptions: $-\big(m''-
  \frac{3}{2}\frac{m'^2}{m}\big)/m^2$(magenta), $\big(m''-2\frac{m'^2}{m}\big)/m^2$(cyan), $-\frac{1}{16}\kappa^2/2$(black), see
  Eqs.(\ref{eq:H1red}), (\ref{eq:H1-H2}) and (\ref{eq:geo}).
 \label{fig:helixmasb}}
\end{figure}

The first and second order derivatives of the mass are then easily
calculated.  We show their angular dependence in
figs.~\ref{fig:helixmasa} and \ref{fig:helixmasb} and as well the
expressions in Eqs.(\ref{eq:H1red}), (\ref{eq:H1-H2}) and (\ref{eq:geo})
that enters the expressions for the different quantizations.  The two
parameterizations can be viewed as stretching and squeezing,
respectively.  The structure variations in these quantities are
therefore rather similar, except that they appear at small or large
$\phi$, respectively as seen in figs.~\ref{fig:helixmasa} and
\ref{fig:helixmasb}.

The mass itself increases quadratically or decreases inversely
proportional with $\phi$.  This smooth dependence is then the origin
of the effects of this type of non-constant mass on the quantization.
The different combinations are then very trivial as for example both
$m''$ and $m'^2/m$ are constants. The difference between $H_{EM_1}$ and
$H_{EM_2}$ therefore quickly vanishes with $\phi$ in the first case while
by construction identically equal to zero in the last case.  The
different combinations exhibit the opposite behavior with $\phi$.  The
additional potential in Eq.(\ref{eq:fullHgeo}) is very small for both
parameterizations.  On the other hand, the geometric potential is again
very decisive with prominent minima at either small or large values of
$\phi$.

\begin{table*}
 \caption{The lowest four eigenvalues for different hamiltonians
   describing one particle on a wire parametrized in
   Eq.~(\ref{eq:phisquared}) with $a=0.1$ and $\phi$ varying between
   $\phi_{min}=0$ and $\phi_{max}=20$.  The ground state energies are in
   absolute values in units of $\hbar^2/(R^2 m_0)$.  The excited
   states are reported as excitation energies related to the
   corresponding ground state energy.
 \label{tab:phisquared}}
 \begin{ruledtabular}
 \begin{tabular}{c| c c c c c}
Hamiltonian		&	$H_{EM_1}$			&	$H_{EM_2}$ 			& JWKB 				&	$H_{geo}$		 &JWKB (including $V_{geo}$)\\
\hline
Ground state	&$2.72\cdot10^{-3}$	&$1.57\cdot10^{-3}$	&$2.29\cdot10^{-3}$ &$-1.51\cdot10^{-2}$ &$-6.01\cdot10^{-3}$ \\
\hline
1st excited		&$7.00\cdot10^{-3}$	&$6.59\cdot10^{-2}$	&$6.85\cdot10^{-3}$ &$1.69\cdot10^{-2}$ & $7.46\cdot10^{-3}$\\
2nd excited		&$1.84\cdot10^{-2}$	&$7.72\cdot10^{-2}$	&$1.84\cdot10^{-2}$ &$2.63\cdot10^{-2}$ & $1.70\cdot10^{-2}$\\ 
3rd excited		&$3.43\cdot10^{-2}$ &$9.31\cdot10^{-2}$	&$3.44\cdot10^{-2}$ &$4.11\cdot10^{-2}$ &$3.19\cdot10^{-2}$\\
\end{tabular}

 \end{ruledtabular}
\end{table*}

\begin{table*}
 \caption{The same as table \ref{tab:phisquared} for the
   parametrization Eq.~(\ref{eq:inversedef}) with $c_0=c_1=0.01$ and
   $\phi$ varying between $\phi_{min}=0$ and $\phi_{max}=40$.
\label{tab:inversedefenergies}}
 \begin{ruledtabular}
 \begin{tabular}{c| c c c c c}
Hamiltonian		&	$H_{EM_1}$			&	$H_{EM_2}$ 			& JWKB 				&	$H_{geo}$		 &JWKB (including $V_{geo}$)\\
\hline
Ground state	&$3.93\cdot10^{-4}$	&$3.93\cdot10^{-4}$	&$4.22\cdot10^{-4}$ &$-8.99\cdot10^{-3}$ &$-6.99\cdot10^{-3}$ \\
\hline
1st excited		&$1.26\cdot10^{-3}$	&$1.26\cdot10^{-3}$	&$1.27\cdot10^{-3}$ &$6.23\cdot10^{-3}$ & $5.25\cdot10^{-3}$\\
2nd excited		&$3.37\cdot10^{-3}$	&$3.37\cdot10^{-3}$	&$3.38\cdot10^{-3}$ &$9.23\cdot10^{-3}$ & $7.38\cdot10^{-3}$\\ 
3rd excited		&$6.32\cdot10^{-3}$ &$6.32\cdot10^{-3}$	&$6.34\cdot10^{-3}$ &$1.17\cdot10^{-2}$ &$9.77\cdot10^{-3}$\\
\end{tabular}

 \end{ruledtabular}
\end{table*}

\subsection{Spectra}

In tables \ref{tab:phisquared} and \ref{tab:inversedefenergies} we
show the lowest four states of the excitation spectra of each
hamiltonian.  The $H_{EM_1}$ and $H_{EM_2}$ energies for both parameterizations
are very close to the corresponding JWKB spectra where the differences
almost disappear for the highest excited states.  The deviation is
largest between the absolute values of the ground state energies of
the $H_{EM_1}$ and $H_{EM_2}$ hamiltonians.  The largest differences between the
parameterizations can be removed by the length scaling which is cleanly
expressed by the analytic JWKB expression in Eq.~(\ref{wkbenergy}).

Inclusion of the geometric potential changes the spectra
substantially.  This potential is attractive and able to support one
or two bound state with negative energy, respectively for the two
parameterizations.  These features are also found in the corresponding
JWKB spectra.  Furthermore, the JWKB excitation energies are approached
for higher excitations.  The length scaling is not appropriate here,
since the wave functions are pulled into the attraction regions which
eliminate the importance of the finite size confinement.

\subsection{Eigenfunctions of stretched helix}

\begin{figure}
\centering
\begingroup
  \makeatletter
  \providecommand\color[2][]{%
    \GenericError{(gnuplot) \space\space\space\@spaces}{%
      Package color not loaded in conjunction with
      terminal option `colourtext'%
    }{See the gnuplot documentation for explanation.%
    }{Either use 'blacktext' in gnuplot or load the package
      color.sty in LaTeX.}%
    \renewcommand\color[2][]{}%
  }%
  \providecommand\includegraphics[2][]{%
    \GenericError{(gnuplot) \space\space\space\@spaces}{%
      Package graphicx or graphics not loaded%
    }{See the gnuplot documentation for explanation.%
    }{The gnuplot epslatex terminal needs graphicx.sty or graphics.sty.}%
    \renewcommand\includegraphics[2][]{}%
  }%
  \providecommand\rotatebox[2]{#2}%
  \@ifundefined{ifGPcolor}{%
    \newif\ifGPcolor
    \GPcolortrue
  }{}%
  \@ifundefined{ifGPblacktext}{%
    \newif\ifGPblacktext
    \GPblacktextfalse
  }{}%
  \let\gplgaddtomacro\g@addto@macro
  \gdef\gplbacktext{}%
  \gdef\gplfronttext{}%
  \makeatother
  \ifGPblacktext
    \def\colorrgb#1{}%
    \def\colorgray#1{}%
  \else
    \ifGPcolor
      \def\colorrgb#1{\color[rgb]{#1}}%
      \def\colorgray#1{\color[gray]{#1}}%
      \expandafter\def\csname LTw\endcsname{\color{white}}%
      \expandafter\def\csname LTb\endcsname{\color{black}}%
      \expandafter\def\csname LTa\endcsname{\color{black}}%
      \expandafter\def\csname LT0\endcsname{\color[rgb]{1,0,0}}%
      \expandafter\def\csname LT1\endcsname{\color[rgb]{0,1,0}}%
      \expandafter\def\csname LT2\endcsname{\color[rgb]{0,0,1}}%
      \expandafter\def\csname LT3\endcsname{\color[rgb]{1,0,1}}%
      \expandafter\def\csname LT4\endcsname{\color[rgb]{0,1,1}}%
      \expandafter\def\csname LT5\endcsname{\color[rgb]{1,1,0}}%
      \expandafter\def\csname LT6\endcsname{\color[rgb]{0,0,0}}%
      \expandafter\def\csname LT7\endcsname{\color[rgb]{1,0.3,0}}%
      \expandafter\def\csname LT8\endcsname{\color[rgb]{0.5,0.5,0.5}}%
    \else
      \def\colorrgb#1{\color{black}}%
      \def\colorgray#1{\color[gray]{#1}}%
      \expandafter\def\csname LTw\endcsname{\color{white}}%
      \expandafter\def\csname LTb\endcsname{\color{black}}%
      \expandafter\def\csname LTa\endcsname{\color{black}}%
      \expandafter\def\csname LT0\endcsname{\color{black}}%
      \expandafter\def\csname LT1\endcsname{\color{black}}%
      \expandafter\def\csname LT2\endcsname{\color{black}}%
      \expandafter\def\csname LT3\endcsname{\color{black}}%
      \expandafter\def\csname LT4\endcsname{\color{black}}%
      \expandafter\def\csname LT5\endcsname{\color{black}}%
      \expandafter\def\csname LT6\endcsname{\color{black}}%
      \expandafter\def\csname LT7\endcsname{\color{black}}%
      \expandafter\def\csname LT8\endcsname{\color{black}}%
    \fi
  \fi
  \setlength{\unitlength}{0.0500bp}%
  \begin{picture}(4676.00,3968.00)%
    \gplgaddtomacro\gplbacktext{%
      \csname LTb\endcsname%
      \put(946,704){\makebox(0,0)[r]{\strut{} 0}}%
      \put(946,1204){\makebox(0,0)[r]{\strut{} 0.1}}%
      \put(946,1704){\makebox(0,0)[r]{\strut{} 0.2}}%
      \put(946,2204){\makebox(0,0)[r]{\strut{} 0.3}}%
      \put(946,2703){\makebox(0,0)[r]{\strut{} 0.4}}%
      \put(946,3203){\makebox(0,0)[r]{\strut{} 0.5}}%
      \put(946,3703){\makebox(0,0)[r]{\strut{} 0.6}}%
      \put(1078,484){\makebox(0,0){\strut{} 0}}%
      \put(1878,484){\makebox(0,0){\strut{} 5}}%
      \put(2679,484){\makebox(0,0){\strut{} 10}}%
      \put(3479,484){\makebox(0,0){\strut{} 15}}%
      \put(4279,484){\makebox(0,0){\strut{} 20}}%
      \put(176,2203){\rotatebox{-270}{\makebox(0,0){\strut{}$\psi(\phi)$}}}%
      \put(2678,154){\makebox(0,0){\strut{}$\phi$}}%
    }%
    \gplgaddtomacro\gplfronttext{%
      \csname LTb\endcsname%
      \put(1738,3530){\makebox(0,0)[r]{\strut{}$H_{geo}$}}%
      \csname LTb\endcsname%
      \put(1738,3310){\makebox(0,0)[r]{\strut{}$H_{EM_2}$}}%
      \csname LTb\endcsname%
      \put(3121,3530){\makebox(0,0)[r]{\strut{}$H_{EM_1}$}}%
      \csname LTb\endcsname%
      \put(3121,3310){\makebox(0,0)[r]{\strut{}JWKB}}%
    }%
    \gplgaddtomacro\gplbacktext{%
      \csname LTb\endcsname%
      \put(946,704){\makebox(0,0)[r]{\strut{} 0}}%
      \put(946,1204){\makebox(0,0)[r]{\strut{} 0.1}}%
      \put(946,1704){\makebox(0,0)[r]{\strut{} 0.2}}%
      \put(946,2204){\makebox(0,0)[r]{\strut{} 0.3}}%
      \put(946,2703){\makebox(0,0)[r]{\strut{} 0.4}}%
      \put(946,3203){\makebox(0,0)[r]{\strut{} 0.5}}%
      \put(946,3703){\makebox(0,0)[r]{\strut{} 0.6}}%
      \put(1078,484){\makebox(0,0){\strut{} 0}}%
      \put(1878,484){\makebox(0,0){\strut{} 5}}%
      \put(2679,484){\makebox(0,0){\strut{} 10}}%
      \put(3479,484){\makebox(0,0){\strut{} 15}}%
      \put(4279,484){\makebox(0,0){\strut{} 20}}%
      \put(176,2203){\rotatebox{-270}{\makebox(0,0){\strut{}$\psi(\phi)$}}}%
      \put(2678,154){\makebox(0,0){\strut{}$\phi$}}%
    }%
    \gplgaddtomacro\gplfronttext{%
      \csname LTb\endcsname%
      \put(2266,3093){\makebox(0,0)[r]{\strut{}JWKB $+V_{geo}$}}%
    }%
    \gplbacktext
    \put(0,0){\includegraphics{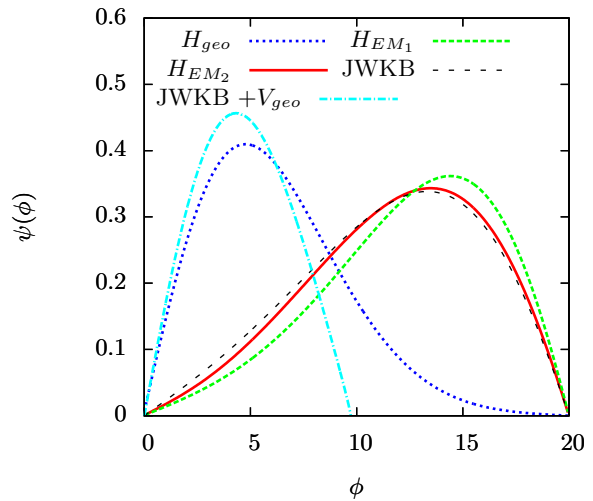}}%
    \gplfronttext
  \end{picture}%
\endgroup
\caption{Ground states for the different choices of hamiltonian,
  Eq.(\ref{eq:H2}) (red), Eq.((\ref{eq:pmp}) (green),
  Eq.(\ref{wkbwave}) (black) for the parametrization in
  Eq.(\ref{eq:phisquared}). The constants in the parametrization are
  chosen to be $a=0.1$.
 \label{fig:phia01n1}}
\end{figure}

The stretched helix is parametrized by Eq.~(\ref{eq:phisquared}) where
we choose again the stretching parameter $a$ to have the value $a=0.1$.
The parametric angle interval for the curve is $\phi$ varying between
$\phi_{min}=0$ and $\phi_{max}=20$.  The ground state wavefunctions for
all the different hamiltonians are shown in fig.~\ref{fig:phia01n1}.
The solution to $H_{EM_1}$ and $H_{EM_2}$ and the corresponding JWKB result all
exhibit a single maximum shifted from the center at $\phi=10$ towards
higher values of $\phi$.  These wave functions are very similar,
although that of $H_{EM_1}$ has the maximum shifted a little more than the
almost indistinguishable results for $H_{EM_2}$ and the JWKB approximation.
These shifts are towards higher values of the effective mass and all
due to the corresponding increase with $\phi$.

The geometric potential has a very strong effect. The ground state
wavefunctions still only have one peak, but now shifted towards
smaller values of $\phi$, where the curvature is larger and the
attraction therefore stronger.  The corresponding JWKB solution is
similar with one peak at roughly the same position as the full
solution.  However, the classically forbidden region of
$V_{geo}(\phi)>E$ starts at $\phi = 10$, after which the JWKB
wavefunction is zero by definition.

\begin{figure}
\centering
\begingroup
  \makeatletter
  \providecommand\color[2][]{%
    \GenericError{(gnuplot) \space\space\space\@spaces}{%
      Package color not loaded in conjunction with
      terminal option `colourtext'%
    }{See the gnuplot documentation for explanation.%
    }{Either use 'blacktext' in gnuplot or load the package
      color.sty in LaTeX.}%
    \renewcommand\color[2][]{}%
  }%
  \providecommand\includegraphics[2][]{%
    \GenericError{(gnuplot) \space\space\space\@spaces}{%
      Package graphicx or graphics not loaded%
    }{See the gnuplot documentation for explanation.%
    }{The gnuplot epslatex terminal needs graphicx.sty or graphics.sty.}%
    \renewcommand\includegraphics[2][]{}%
  }%
  \providecommand\rotatebox[2]{#2}%
  \@ifundefined{ifGPcolor}{%
    \newif\ifGPcolor
    \GPcolortrue
  }{}%
  \@ifundefined{ifGPblacktext}{%
    \newif\ifGPblacktext
    \GPblacktextfalse
  }{}%
  \let\gplgaddtomacro\g@addto@macro
  \gdef\gplbacktext{}%
  \gdef\gplfronttext{}%
  \makeatother
  \ifGPblacktext
    \def\colorrgb#1{}%
    \def\colorgray#1{}%
  \else
    \ifGPcolor
      \def\colorrgb#1{\color[rgb]{#1}}%
      \def\colorgray#1{\color[gray]{#1}}%
      \expandafter\def\csname LTw\endcsname{\color{white}}%
      \expandafter\def\csname LTb\endcsname{\color{black}}%
      \expandafter\def\csname LTa\endcsname{\color{black}}%
      \expandafter\def\csname LT0\endcsname{\color[rgb]{1,0,0}}%
      \expandafter\def\csname LT1\endcsname{\color[rgb]{0,1,0}}%
      \expandafter\def\csname LT2\endcsname{\color[rgb]{0,0,1}}%
      \expandafter\def\csname LT3\endcsname{\color[rgb]{1,0,1}}%
      \expandafter\def\csname LT4\endcsname{\color[rgb]{0,1,1}}%
      \expandafter\def\csname LT5\endcsname{\color[rgb]{1,1,0}}%
      \expandafter\def\csname LT6\endcsname{\color[rgb]{0,0,0}}%
      \expandafter\def\csname LT7\endcsname{\color[rgb]{1,0.3,0}}%
      \expandafter\def\csname LT8\endcsname{\color[rgb]{0.5,0.5,0.5}}%
    \else
      \def\colorrgb#1{\color{black}}%
      \def\colorgray#1{\color[gray]{#1}}%
      \expandafter\def\csname LTw\endcsname{\color{white}}%
      \expandafter\def\csname LTb\endcsname{\color{black}}%
      \expandafter\def\csname LTa\endcsname{\color{black}}%
      \expandafter\def\csname LT0\endcsname{\color{black}}%
      \expandafter\def\csname LT1\endcsname{\color{black}}%
      \expandafter\def\csname LT2\endcsname{\color{black}}%
      \expandafter\def\csname LT3\endcsname{\color{black}}%
      \expandafter\def\csname LT4\endcsname{\color{black}}%
      \expandafter\def\csname LT5\endcsname{\color{black}}%
      \expandafter\def\csname LT6\endcsname{\color{black}}%
      \expandafter\def\csname LT7\endcsname{\color{black}}%
      \expandafter\def\csname LT8\endcsname{\color{black}}%
    \fi
  \fi
  \setlength{\unitlength}{0.0500bp}%
  \begin{picture}(4676.00,3968.00)%
    \gplgaddtomacro\gplbacktext{%
      \csname LTb\endcsname%
      \put(946,977){\makebox(0,0)[r]{\strut{}-0.4}}%
      \put(946,1522){\makebox(0,0)[r]{\strut{}-0.2}}%
      \put(946,2067){\makebox(0,0)[r]{\strut{} 0}}%
      \put(946,2612){\makebox(0,0)[r]{\strut{} 0.2}}%
      \put(946,3158){\makebox(0,0)[r]{\strut{} 0.4}}%
      \put(946,3703){\makebox(0,0)[r]{\strut{} 0.6}}%
      \put(1078,484){\makebox(0,0){\strut{} 0}}%
      \put(1878,484){\makebox(0,0){\strut{} 5}}%
      \put(2679,484){\makebox(0,0){\strut{} 10}}%
      \put(3479,484){\makebox(0,0){\strut{} 15}}%
      \put(4279,484){\makebox(0,0){\strut{} 20}}%
      \put(176,2203){\rotatebox{-270}{\makebox(0,0){\strut{}$\psi(\phi)$}}}%
      \put(2678,154){\makebox(0,0){\strut{}$\phi$}}%
    }%
    \gplgaddtomacro\gplfronttext{%
      \csname LTb\endcsname%
      \put(1738,3530){\makebox(0,0)[r]{\strut{}$H_{geo}$}}%
      \csname LTb\endcsname%
      \put(1738,3310){\makebox(0,0)[r]{\strut{}$H_{EM_2}$}}%
      \csname LTb\endcsname%
      \put(3121,3530){\makebox(0,0)[r]{\strut{}$H_{EM_1}$}}%
      \csname LTb\endcsname%
      \put(3121,3310){\makebox(0,0)[r]{\strut{}JWKB}}%
    }%
    \gplgaddtomacro\gplbacktext{%
      \csname LTb\endcsname%
      \put(946,977){\makebox(0,0)[r]{\strut{}-0.4}}%
      \put(946,1522){\makebox(0,0)[r]{\strut{}-0.2}}%
      \put(946,2067){\makebox(0,0)[r]{\strut{} 0}}%
      \put(946,2612){\makebox(0,0)[r]{\strut{} 0.2}}%
      \put(946,3158){\makebox(0,0)[r]{\strut{} 0.4}}%
      \put(946,3703){\makebox(0,0)[r]{\strut{} 0.6}}%
      \put(1078,484){\makebox(0,0){\strut{} 0}}%
      \put(1878,484){\makebox(0,0){\strut{} 5}}%
      \put(2679,484){\makebox(0,0){\strut{} 10}}%
      \put(3479,484){\makebox(0,0){\strut{} 15}}%
      \put(4279,484){\makebox(0,0){\strut{} 20}}%
      \put(176,2203){\rotatebox{-270}{\makebox(0,0){\strut{}$\psi(\phi)$}}}%
      \put(2678,154){\makebox(0,0){\strut{}$\phi$}}%
    }%
    \gplgaddtomacro\gplfronttext{%
      \csname LTb\endcsname%
      \put(2266,3048){\makebox(0,0)[r]{\strut{}JWKB $+V_{geo}$}}%
    }%
    \gplbacktext
    \put(0,0){\includegraphics{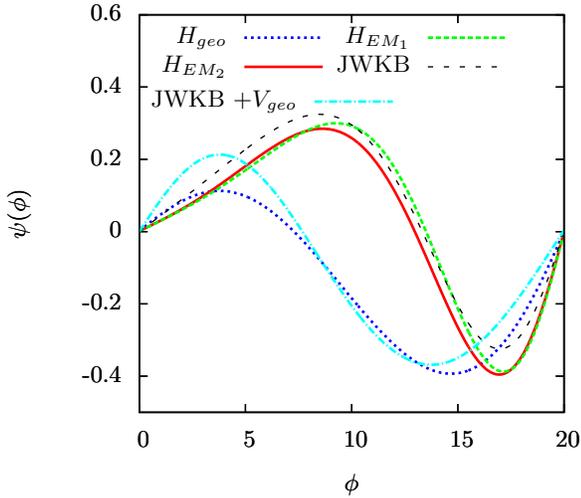}}%
    \gplfronttext
  \end{picture}%
\endgroup
\caption{ Excited states for the different choices of hamiltonian,
  Eq.(\ref{eq:H2}) (red), Eq.((\ref{eq:pmp}) (green),
  Eq.(\ref{wkbwave}) (black) for the parametrization in
  Eq.(\ref{eq:phisquared}).  The constants in the parametrization are
  chosen to be $a=0.1$.
 \label{fig:phia01n2}}
\end{figure}

The wavefunctions for the first excited state are shown in
fig.~\ref{fig:phia01n2}, where the necessary node is the prominent
feature.  Again almost quantitative agreement within the two groups of
results, that is between the $H_{EM_1}$, $H_{EM_2}$ and related JWKB results, and
between the $H_{geo}$ and corresponding JWKB results.  The added
geometric potential changes the quantitative behavior by moving the
peaks towards smaller $\phi$-values where the attraction is largest.

\subsection{Eigenfunctions of squeezed helix}

\begin{figure}
\centering
\begingroup
  \makeatletter
  \providecommand\color[2][]{%
    \GenericError{(gnuplot) \space\space\space\@spaces}{%
      Package color not loaded in conjunction with
      terminal option `colourtext'%
    }{See the gnuplot documentation for explanation.%
    }{Either use 'blacktext' in gnuplot or load the package
      color.sty in LaTeX.}%
    \renewcommand\color[2][]{}%
  }%
  \providecommand\includegraphics[2][]{%
    \GenericError{(gnuplot) \space\space\space\@spaces}{%
      Package graphicx or graphics not loaded%
    }{See the gnuplot documentation for explanation.%
    }{The gnuplot epslatex terminal needs graphicx.sty or graphics.sty.}%
    \renewcommand\includegraphics[2][]{}%
  }%
  \providecommand\rotatebox[2]{#2}%
  \@ifundefined{ifGPcolor}{%
    \newif\ifGPcolor
    \GPcolortrue
  }{}%
  \@ifundefined{ifGPblacktext}{%
    \newif\ifGPblacktext
    \GPblacktextfalse
  }{}%
  \let\gplgaddtomacro\g@addto@macro
  \gdef\gplbacktext{}%
  \gdef\gplfronttext{}%
  \makeatother
  \ifGPblacktext
    \def\colorrgb#1{}%
    \def\colorgray#1{}%
  \else
    \ifGPcolor
      \def\colorrgb#1{\color[rgb]{#1}}%
      \def\colorgray#1{\color[gray]{#1}}%
      \expandafter\def\csname LTw\endcsname{\color{white}}%
      \expandafter\def\csname LTb\endcsname{\color{black}}%
      \expandafter\def\csname LTa\endcsname{\color{black}}%
      \expandafter\def\csname LT0\endcsname{\color[rgb]{1,0,0}}%
      \expandafter\def\csname LT1\endcsname{\color[rgb]{0,1,0}}%
      \expandafter\def\csname LT2\endcsname{\color[rgb]{0,0,1}}%
      \expandafter\def\csname LT3\endcsname{\color[rgb]{1,0,1}}%
      \expandafter\def\csname LT4\endcsname{\color[rgb]{0,1,1}}%
      \expandafter\def\csname LT5\endcsname{\color[rgb]{1,1,0}}%
      \expandafter\def\csname LT6\endcsname{\color[rgb]{0,0,0}}%
      \expandafter\def\csname LT7\endcsname{\color[rgb]{1,0.3,0}}%
      \expandafter\def\csname LT8\endcsname{\color[rgb]{0.5,0.5,0.5}}%
    \else
      \def\colorrgb#1{\color{black}}%
      \def\colorgray#1{\color[gray]{#1}}%
      \expandafter\def\csname LTw\endcsname{\color{white}}%
      \expandafter\def\csname LTb\endcsname{\color{black}}%
      \expandafter\def\csname LTa\endcsname{\color{black}}%
      \expandafter\def\csname LT0\endcsname{\color{black}}%
      \expandafter\def\csname LT1\endcsname{\color{black}}%
      \expandafter\def\csname LT2\endcsname{\color{black}}%
      \expandafter\def\csname LT3\endcsname{\color{black}}%
      \expandafter\def\csname LT4\endcsname{\color{black}}%
      \expandafter\def\csname LT5\endcsname{\color{black}}%
      \expandafter\def\csname LT6\endcsname{\color{black}}%
      \expandafter\def\csname LT7\endcsname{\color{black}}%
      \expandafter\def\csname LT8\endcsname{\color{black}}%
    \fi
  \fi
  \setlength{\unitlength}{0.0500bp}%
  \begin{picture}(4534.00,3968.00)%
    \gplgaddtomacro\gplbacktext{%
      \csname LTb\endcsname%
      \put(1078,704){\makebox(0,0)[r]{\strut{} 0}}%
      \put(1078,1037){\makebox(0,0)[r]{\strut{} 0.05}}%
      \put(1078,1370){\makebox(0,0)[r]{\strut{} 0.10}}%
      \put(1078,1704){\makebox(0,0)[r]{\strut{} 0.15}}%
      \put(1078,2037){\makebox(0,0)[r]{\strut{} 0.20}}%
      \put(1078,2370){\makebox(0,0)[r]{\strut{} 0.25}}%
      \put(1078,2703){\makebox(0,0)[r]{\strut{} 0.30}}%
      \put(1078,3037){\makebox(0,0)[r]{\strut{} 0.35}}%
      \put(1078,3370){\makebox(0,0)[r]{\strut{} 0.40}}%
      \put(1078,3703){\makebox(0,0)[r]{\strut{} 0.45}}%
      \put(1210,484){\makebox(0,0){\strut{} 0}}%
      \put(1576,484){\makebox(0,0){\strut{} 5}}%
      \put(1942,484){\makebox(0,0){\strut{} 10}}%
      \put(2308,484){\makebox(0,0){\strut{} 15}}%
      \put(2674,484){\makebox(0,0){\strut{} 20}}%
      \put(3039,484){\makebox(0,0){\strut{} 25}}%
      \put(3405,484){\makebox(0,0){\strut{} 30}}%
      \put(3771,484){\makebox(0,0){\strut{} 35}}%
      \put(4137,484){\makebox(0,0){\strut{} 40}}%
      \put(176,2203){\rotatebox{-270}{\makebox(0,0){\strut{}$\psi(\phi)$}}}%
      \put(2673,154){\makebox(0,0){\strut{}$\phi$}}%
    }%
    \gplgaddtomacro\gplfronttext{%
      \csname LTb\endcsname%
      \put(1870,3530){\makebox(0,0)[r]{\strut{}$H_{geo}$}}%
      \csname LTb\endcsname%
      \put(1870,3310){\makebox(0,0)[r]{\strut{}$H_{EM_2}$}}%
      \csname LTb\endcsname%
      \put(3253,3530){\makebox(0,0)[r]{\strut{}$H_{EM_1}$}}%
      \csname LTb\endcsname%
      \put(3253,3310){\makebox(0,0)[r]{\strut{}JWKB}}%
    }%
    \gplgaddtomacro\gplbacktext{%
      \csname LTb\endcsname%
      \put(1078,704){\makebox(0,0)[r]{\strut{} 0}}%
      \put(1078,1037){\makebox(0,0)[r]{\strut{} 0.05}}%
      \put(1078,1370){\makebox(0,0)[r]{\strut{} 0.10}}%
      \put(1078,1704){\makebox(0,0)[r]{\strut{} 0.15}}%
      \put(1078,2037){\makebox(0,0)[r]{\strut{} 0.20}}%
      \put(1078,2370){\makebox(0,0)[r]{\strut{} 0.25}}%
      \put(1078,2703){\makebox(0,0)[r]{\strut{} 0.30}}%
      \put(1078,3037){\makebox(0,0)[r]{\strut{} 0.35}}%
      \put(1078,3370){\makebox(0,0)[r]{\strut{} 0.40}}%
      \put(1078,3703){\makebox(0,0)[r]{\strut{} 0.45}}%
      \put(1210,484){\makebox(0,0){\strut{} 0}}%
      \put(1576,484){\makebox(0,0){\strut{} 5}}%
      \put(1942,484){\makebox(0,0){\strut{} 10}}%
      \put(2308,484){\makebox(0,0){\strut{} 15}}%
      \put(2674,484){\makebox(0,0){\strut{} 20}}%
      \put(3039,484){\makebox(0,0){\strut{} 25}}%
      \put(3405,484){\makebox(0,0){\strut{} 30}}%
      \put(3771,484){\makebox(0,0){\strut{} 35}}%
      \put(4137,484){\makebox(0,0){\strut{} 40}}%
      \put(176,2203){\rotatebox{-270}{\makebox(0,0){\strut{}$\psi(\phi)$}}}%
      \put(2673,154){\makebox(0,0){\strut{}$\phi$}}%
    }%
    \gplgaddtomacro\gplfronttext{%
      \csname LTb\endcsname%
      \put(2398,3060){\makebox(0,0)[r]{\strut{}JWKB $+V_{geo}$}}%
    }%
    \gplbacktext
    \put(0,0){\includegraphics{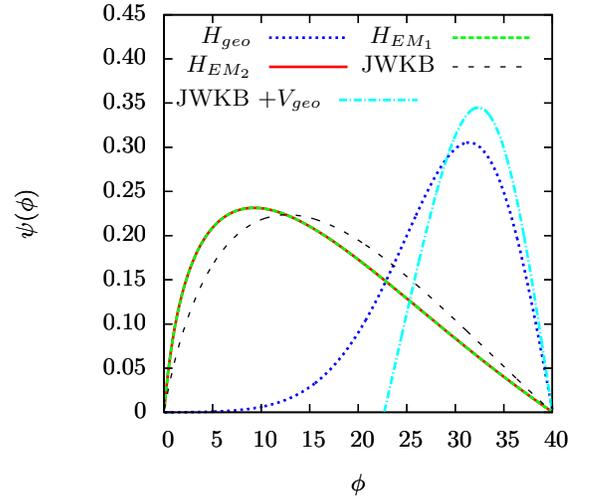}}%
    \gplfronttext
  \end{picture}%
\endgroup
\caption{Ground states for the different choices of hamiltonian,
  Eq.(\ref{eq:H2}) (red), Eq.((\ref{eq:pmp}) (green),
  Eq.(\ref{wkbwave}) (black) for the parametrization in
  Eq.(\ref{eq:phisquared}).  The constants in the parametrization are
  chosen to be $c_0=0.01$ and $c_1=0.01$
\label{fig:inversec0001c1001}.}
\end{figure}

The squeezed helix, which is parameterized by
Eq.~(\ref{eq:inversedef}), is designed to give $H_{EM_1}=H_{EM_2}$.  The
effective mass decreases with $\phi$ as
$m(\phi)=\frac{1}{c_0+c_1\phi}$, where we choose $c_0=c_1=0.01$ and a
curve parametrized by $\phi$ varying between $\phi_{min}=0$ and
$\phi_{max}=40$.  We show the ground state wavefunctions in
fig.~\ref{fig:inversec0001c1001}.  They are almost a left-right
reflection of the stretched wave functions in fig.~\ref{fig:phia01n1}.
The overlapping wavefunctions of $H_{EM_1}$ and $H_{EM_2}$ rise quickly from
zero to a single maximum at $\phi=9$, before they linearly fall off to
$0$ at $\phi_{max}$.  The JWKB solution is similar and rises from $0$
to a maximum at $\phi=12$, before it falls off to zero at $\phi_{max}$.

Again the geometric potential moves the peak to the opposite end of the
allowed $\phi$ interval, that is to about $\phi=32$.  The
corresponding JWKB solution is similar with a maximum at roughly the
same $\phi$ value. The decrease is steeper towards the classically
forbidden region for $\phi<22$.  Thus the picture is that the
geometric potential move the solutions to the large curvature region,
which is the opposite of the large effective mass region where the
$H_{EM_1}$ and $H_{EM_2}$ solutions are peaked.

\begin{figure}
\centering
\begingroup
  \makeatletter
  \providecommand\color[2][]{%
    \GenericError{(gnuplot) \space\space\space\@spaces}{%
      Package color not loaded in conjunction with
      terminal option `colourtext'%
    }{See the gnuplot documentation for explanation.%
    }{Either use 'blacktext' in gnuplot or load the package
      color.sty in LaTeX.}%
    \renewcommand\color[2][]{}%
  }%
  \providecommand\includegraphics[2][]{%
    \GenericError{(gnuplot) \space\space\space\@spaces}{%
      Package graphicx or graphics not loaded%
    }{See the gnuplot documentation for explanation.%
    }{The gnuplot epslatex terminal needs graphicx.sty or graphics.sty.}%
    \renewcommand\includegraphics[2][]{}%
  }%
  \providecommand\rotatebox[2]{#2}%
  \@ifundefined{ifGPcolor}{%
    \newif\ifGPcolor
    \GPcolortrue
  }{}%
  \@ifundefined{ifGPblacktext}{%
    \newif\ifGPblacktext
    \GPblacktextfalse
  }{}%
  \let\gplgaddtomacro\g@addto@macro
  \gdef\gplbacktext{}%
  \gdef\gplfronttext{}%
  \makeatother
  \ifGPblacktext
    \def\colorrgb#1{}%
    \def\colorgray#1{}%
  \else
    \ifGPcolor
      \def\colorrgb#1{\color[rgb]{#1}}%
      \def\colorgray#1{\color[gray]{#1}}%
      \expandafter\def\csname LTw\endcsname{\color{white}}%
      \expandafter\def\csname LTb\endcsname{\color{black}}%
      \expandafter\def\csname LTa\endcsname{\color{black}}%
      \expandafter\def\csname LT0\endcsname{\color[rgb]{1,0,0}}%
      \expandafter\def\csname LT1\endcsname{\color[rgb]{0,1,0}}%
      \expandafter\def\csname LT2\endcsname{\color[rgb]{0,0,1}}%
      \expandafter\def\csname LT3\endcsname{\color[rgb]{1,0,1}}%
      \expandafter\def\csname LT4\endcsname{\color[rgb]{0,1,1}}%
      \expandafter\def\csname LT5\endcsname{\color[rgb]{1,1,0}}%
      \expandafter\def\csname LT6\endcsname{\color[rgb]{0,0,0}}%
      \expandafter\def\csname LT7\endcsname{\color[rgb]{1,0.3,0}}%
      \expandafter\def\csname LT8\endcsname{\color[rgb]{0.5,0.5,0.5}}%
    \else
      \def\colorrgb#1{\color{black}}%
      \def\colorgray#1{\color[gray]{#1}}%
      \expandafter\def\csname LTw\endcsname{\color{white}}%
      \expandafter\def\csname LTb\endcsname{\color{black}}%
      \expandafter\def\csname LTa\endcsname{\color{black}}%
      \expandafter\def\csname LT0\endcsname{\color{black}}%
      \expandafter\def\csname LT1\endcsname{\color{black}}%
      \expandafter\def\csname LT2\endcsname{\color{black}}%
      \expandafter\def\csname LT3\endcsname{\color{black}}%
      \expandafter\def\csname LT4\endcsname{\color{black}}%
      \expandafter\def\csname LT5\endcsname{\color{black}}%
      \expandafter\def\csname LT6\endcsname{\color{black}}%
      \expandafter\def\csname LT7\endcsname{\color{black}}%
      \expandafter\def\csname LT8\endcsname{\color{black}}%
    \fi
  \fi
  \setlength{\unitlength}{0.0500bp}%
  \begin{picture}(4534.00,3968.00)%
    \gplgaddtomacro\gplbacktext{%
      \csname LTb\endcsname%
      \put(946,704){\makebox(0,0)[r]{\strut{}-0.3}}%
      \put(946,1057){\makebox(0,0)[r]{\strut{}-0.2}}%
      \put(946,1410){\makebox(0,0)[r]{\strut{}-0.1}}%
      \put(946,1762){\makebox(0,0)[r]{\strut{} 0}}%
      \put(946,2115){\makebox(0,0)[r]{\strut{} 0.1}}%
      \put(946,2468){\makebox(0,0)[r]{\strut{} 0.2}}%
      \put(946,2821){\makebox(0,0)[r]{\strut{} 0.3}}%
      \put(946,3174){\makebox(0,0)[r]{\strut{} 0.4}}%
      \put(946,3527){\makebox(0,0)[r]{\strut{} 0.5}}%
      \put(1078,484){\makebox(0,0){\strut{} 0}}%
      \put(1460,484){\makebox(0,0){\strut{} 5}}%
      \put(1843,484){\makebox(0,0){\strut{} 10}}%
      \put(2225,484){\makebox(0,0){\strut{} 15}}%
      \put(2608,484){\makebox(0,0){\strut{} 20}}%
      \put(2990,484){\makebox(0,0){\strut{} 25}}%
      \put(3372,484){\makebox(0,0){\strut{} 30}}%
      \put(3755,484){\makebox(0,0){\strut{} 35}}%
      \put(4137,484){\makebox(0,0){\strut{} 40}}%
      \put(176,2203){\rotatebox{-270}{\makebox(0,0){\strut{}$\psi(\phi)$}}}%
      \put(2607,154){\makebox(0,0){\strut{}$\phi$}}%
    }%
    \gplgaddtomacro\gplfronttext{%
      \csname LTb\endcsname%
      \put(1738,3530){\makebox(0,0)[r]{\strut{}$H_{geo}$}}%
      \csname LTb\endcsname%
      \put(1738,3310){\makebox(0,0)[r]{\strut{}$H_{EM_2}$}}%
      \csname LTb\endcsname%
      \put(3121,3530){\makebox(0,0)[r]{\strut{}$H_{EM_1}$}}%
      \csname LTb\endcsname%
      \put(3121,3310){\makebox(0,0)[r]{\strut{}JWKB}}%
    }%
    \gplgaddtomacro\gplbacktext{%
      \csname LTb\endcsname%
      \put(946,704){\makebox(0,0)[r]{\strut{}-0.3}}%
      \put(946,1057){\makebox(0,0)[r]{\strut{}-0.2}}%
      \put(946,1410){\makebox(0,0)[r]{\strut{}-0.1}}%
      \put(946,1762){\makebox(0,0)[r]{\strut{} 0}}%
      \put(946,2115){\makebox(0,0)[r]{\strut{} 0.1}}%
      \put(946,2468){\makebox(0,0)[r]{\strut{} 0.2}}%
      \put(946,2821){\makebox(0,0)[r]{\strut{} 0.3}}%
      \put(946,3174){\makebox(0,0)[r]{\strut{} 0.4}}%
      \put(946,3527){\makebox(0,0)[r]{\strut{} 0.5}}%
      \put(1078,484){\makebox(0,0){\strut{} 0}}%
      \put(1460,484){\makebox(0,0){\strut{} 5}}%
      \put(1843,484){\makebox(0,0){\strut{} 10}}%
      \put(2225,484){\makebox(0,0){\strut{} 15}}%
      \put(2608,484){\makebox(0,0){\strut{} 20}}%
      \put(2990,484){\makebox(0,0){\strut{} 25}}%
      \put(3372,484){\makebox(0,0){\strut{} 30}}%
      \put(3755,484){\makebox(0,0){\strut{} 35}}%
      \put(4137,484){\makebox(0,0){\strut{} 40}}%
      \put(176,2203){\rotatebox{-270}{\makebox(0,0){\strut{}$\psi(\phi)$}}}%
      \put(2607,154){\makebox(0,0){\strut{}$\phi$}}%
    }%
    \gplgaddtomacro\gplfronttext{%
      \csname LTb\endcsname%
      \put(2266,3099){\makebox(0,0)[r]{\strut{}JWKB $+V_{geo}$}}%
    }%
    \gplbacktext
    \put(0,0){\includegraphics{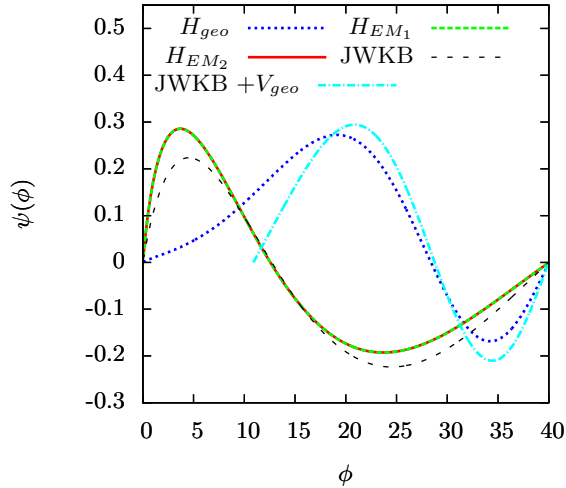}}%
    \gplfronttext
  \end{picture}%
\endgroup
\caption{ First excited states for the different choices of
  hamiltonian, Eq.(\ref{eq:H2}) (red), Eq.((\ref{eq:pmp}) (green),
  Eq.(\ref{wkbwave}) (black) for the parametrization in
  Eq.(\ref{eq:phisquared}).  The constants in the parametrization are
  chosen to be $c_0=0.01$ and $c_1=0.01$
\label{fig:inversec0001c1001n2}.}
\end{figure}

The first excited states of the same configuration are shown in
figure~\ref{fig:inversec0001c1001n2}.  They all now have the required
node for an excited state.  The overlapping solutions from $H_{EM_1}$ and
$H_{EM_2}$ rise to their first maximum at $\phi=4$, then they have a node
at $\phi=11$, and then a smaller minimum at $\phi=23$.  The related
JWKB solutions is similar with slightly shifted extremum points.  The
geometric potential leads to a wavefunction with a broad peak at the
center, and a smaller minimum at the position where the ground state
wavefunction is peaked.  This is required by orthogonality.  The
corresponding JWKB solution is similar but has as usual  to vanish
within the classical forbidden region for $\phi<10$.

\section{Discussion and outlook\label{sec:outlook}}

We start with two different approaches to a system of a single
particle trapped in an effective one dimensional trap.  The first
approach is to build a classical description of the system, and
through that find an appropriate quantization.  Because we allowed
this one dimensional trap to have a changing curvature, this
quantization step is not trivial and we show two equally valid
choices, which differ only by a potential term.

The other approach starts from a quantum mechanical description in
three dimensions, and then through a transverse-mode adiabatic
approximation reduce to an effective one dimensional model but now
with an extra so-called geometric potential.  This potential is
attractive and given as proportional to the square of the curvature.
We then investigate three different perturbations of a helix, and
calculate the wavefunctions and energies of the different
hamiltonians.

Monotonous deformation results in monotonous effective mass and
curvature dependence on the coordinate.  However, these two key
quantities behave differently and lead to opposite effects on the
quantized solutions.  Specifically, increasing curvature leads to
increasing attraction along the wire, and therefore the ground state
wavefunctions would peak at this end.  Increasing effective mass also
tend to move the largest probability in the same direction of large
mass.  This implies that the different quantization prescriptions in
this case of monotonous helix deformation produce very different
results.

The more periodic type of helix deformation leads to more similar
quantized results although still with substantial differences.  The
periodic nature of a helix prohibits that the confining potential is
obtained by a converged Taylor expansion in terms of the parametrizing
one-dimensional path.  It also strongly indicates the same problem
with a quantization obtained by forced, explicit symmetrization of a
non-hermitian hamiltonian.  The only hamiltonian without these
problems has the inverse effective mass between the two derivatives in
the kinetic energy operator. 

The authors acknowledge inspiring conversations with J. Stockhofe
and A. Rauschenbeutel. This work was 
supported by the Danish Council for Independent Research.


\begin{thebibliography}{99}


\bibitem{ricardez2010} Ricardez-Vargas I and Volke-Sep{\'u}lveda K 2010 \emph{J. Opt. Soc. Am.} B {\bf 27} 948 
\bibitem{reitz2012} Reitz D and Rauschenbeutel A 2012 \emph{Opt. Commun.} {\bf 285} 4705 
\bibitem{arnold2012} Arnold A S 2012 \emph{Optics Lett.} {\bf 37} 2505
\bibitem{macdonald2002} MacDonald M P {\it et al} 2002 \emph{Opt. Commun.} {\bf 201} 21 
\bibitem{bhatta2007} Bhattacharya M 2007 \emph{Opt. Commun.} {\bf 279} 219 
\bibitem{sague2008} Sagu{\'e} G, Baade A and Rauschenbeutel A 2008 \emph{New J. Phys.} {\bf 10} 113008 
\bibitem{pang2005} Pang Y K {\it et al} 2005 \emph{Opt. Express} {\bf 13} 7615 
\bibitem{law2008} Law K T and Feldman D E 2008 \emph{Phys. Rev. Lett.} {\bf 101} 096401 
\bibitem{huhta2010} Huhtam{\"a}ki J A M and Kuopanportti P 2010 \emph{Phys. Rev.} A {\bf 82} 053616 
\bibitem{schmelcher2011} Schmelcher P 2011 \emph{Europhys. Lett.} {\bf 95} 50005 
\bibitem{zampetaki2013} Zampetaki A V, Stockhofe J, Kr{\"o}nke S and Schmelcher P 2013 \emph{Phys. Rev.} E {\bf 88} 043202 
\bibitem{pedersen2014} Pedersen J K, Fedorov D V, Jensen A S and Zinner N T 2014 \emph{J. Phys. B:At. Mol. Opt. Phys.} {\bf 47} 165103

\bibitem{sch26} Schr\"odinger E 1926  \emph{Ann. Physik} {\bf 80} 489

\bibitem{hof71} Hofmann H and Dietrich K 1971  \emph{Nucl. Phys.} {\bf A165} 1

\bibitem{pau74} Pauli H C and Ledergerber T 1974
\emph{Proc. Phys. and Chem. of fission} IAEA Rochester, New York, 463

\bibitem{bra72}  Brack M, Damgaard J, Jensen A S, Pauli H C, 
Strutinsky V M and Wong C Y 1972  \emph{Rev. Mod. Phys.} {\bf 44} 320.

\bibitem{PhysRevA.89.033630} Stockhofe J and Schmelcher P 2014
  \emph{Phys. Rev.} A {\bf 89} 033630

\bibitem{taylor2005classical} Taylor J R 2005 \emph{Classical Mechanics} (University Science Books)

\bibitem{dek99} L. Dekar, L. Chetouani, and T.F. Hammann 1999
\emph{Phys. Rev.} A {\bf 59} 107

\bibitem{docarmo1976} do Carmo M P 1976 \emph{Differential Geometry of Curves and Surfaces} (Prentice-Hall)





\end{thebibliography}
\end{document}